\documentclass[prl,amsmath,twocolumn,showpacs]{revtex4}
\usepackage[dvips,pdftex]{graphicx}
\usepackage{hyperref}
\usepackage{amsmath}
\usepackage{amssymb}
\usepackage{mathrsfs}
\usepackage[figuresright]{rotating}
\usepackage{braket}
\usepackage{bm}
\usepackage{amsmath,amssymb}
\usepackage{color}
\usepackage{url}
\usepackage{algorithm}
\usepackage[titletoc]{appendix}
\usepackage{bbm}
\usepackage{bbold}
\usepackage{mathrsfs}
\usepackage[capitalize]{cleveref}
\usepackage{bm,bbm,dsfont}
\usepackage{comment}

\usepackage[utf8]{inputenc}

\begin{document}

\makeatletter
\providecommand{\citenamefont}[1]{#1}
\providecommand{\href}[2]{#2}
\providecommand{\href@noop}[1]{}
\providecommand{\doibase}[1]{https://doi.org/#1}
\providecommand{\natexlab}[1]{#1}
\makeatother


\title{The dynamical law behind eye movements: distinguishing between L\'evy and intermittent strategies}

\author{Pedro Lencastre$^{1,2}$ }
\email{pedroreg@oslomet.no; pedro.lencastre.silva@gmail.com}
\author{Yurii Bystryk$^{3}$}
\author{Anis Yazidi$^{1,2}$}
\author{Sergey Denisov$^{1,2}$}
\email{sergiyde@oslomet.no}
\author{Pedro G.~Lind$^{1,2,4}$}
\email{pedrolin@oslomet.no}

\affiliation{$^1$Department of Computer Science, OsloMet -- Oslo Metropolitan University, N-0130 Oslo, Norway}
\affiliation{$^2${\it AI Lab} -- OsloMet Artificial Intelligence Lab, N-0166 Oslo, Norway}
\affiliation{$^3$Institute of Applied Physics, National Academy of Sciences of Ukraine,  40000 Sumy, Ukraine}
\affiliation{$^4$Simula Research Laboratory, Numerical Analysis and Scientific Computing, 0164 Oslo, Norway}

\begin{abstract}
Foraging is a complex spatio-temporal process which is often described with stochastic models. Two particular ones, L\'evy walks (LWs) and intermittent search (IS), became popular in this context. Researchers from the two communities,  each advocating for either L\'evy or intermittent approach, independently analyzed foraging patterns and 
reported agreement between empirical data and the model they used. We  resolve this  L\'evy-intermittent  dichotomy for eye-gaze trajectories collected in a series of experiments designed to stimulate free foraging for visual information. 
By combining analytical results, statistical quantifiers, and basic machine learning techniques, we devise a method to score the performance of the models when they are used to approximate an individual gaze trajectory. Our analysis indicates that the intermittent search model consistently yields higher scores and thus approximates the majority of the eye-gaze trajectories better. 
\end{abstract}

\maketitle


The trajectories produced by foraging animals often appear to be random-like and, in theoretical ecology,  
stochastic walks are considered one of the most relevant frameworks to model and analyze foraging~\cite{patlak1953random,lin1974mathematics,okubo1980diffusion,kareiva1983analyzing}. Among the multitude of models, 
two have become especially popular over the last two decades:  \emph{L\'evy walks}~\cite{shlesinger1982random,zaburdaev2015levy} (often also refereed to as 'L\'evy flights'~\cite{pyke2015understanding}) and \emph{intermittent search}~\cite{benichou2011intermittent,benichou2006intermittent}.

In its basic formulation, a L\'evy  walk (LW) 
is realized by performing a sequence of relocations,
each carried out in a random direction with a constant velocity $v$~\cite{shlesinger1982random,zaburdaev2015levy}. The duration of each relocation is drawn from a probability density function (PDF) characterized by a power-law tail.
It has been stated that this type of stochastic walks is able to reproduce foraging patterns of living organisms, ranging from chemotactic bacteria ~\cite{ariel2015swarming} and marine predators~\cite{sims2008scaling,humphries2010environmental} to human gatherers~\cite{brown2007levy,raichlen2014evidence,garg2021efficient}, visual foraging~\cite{peinkeEyeT} and even extinct animals~\cite{sims2014hierarchical}. In the case of renewable~\cite{viswanathan1999optimizing} or finite-lived~\cite{boyer2024optimizing} targets, LW 
were advocated as the most optimal foraging strategy, which is considered to be a reason for the universality of the 'L\'evy foraging'~\cite{viswanathan2008levy}. This hypothesis, however, prompted significant criticism, both with respect to the data analysis methodology and theoretical reasoning used to support it~\cite{edwards2007revisiting,edwards2011overturning,reynolds2015liberating,levernier2020inverse,buldyrev2021comment,levernier2021reply}.  

An intermittent search (IS) 
is composed of two alternating phases, a diffusion (\textit{D}), that is 
associated with the searching phase, 
and a ballistic (\textit{B}), associated with relocation~\cite{benichou2005optimal,benichou2006two,benichou2006intermittent,benichou2011intermittent,oshanin2009efficient}. Durations of the phases are governed by the corresponding switching rates, $\lambda_{DB}$ (from \textit{D} to \textit{B}) and $\lambda_{BD}$. 
Experimental data collected on various species have been analyzed and a good agreement with the IS model was 
stated~\cite{benichou2005optimal}. 
\begin{figure}[b]
\begin{center}
\includegraphics[width=0.99\columnwidth]{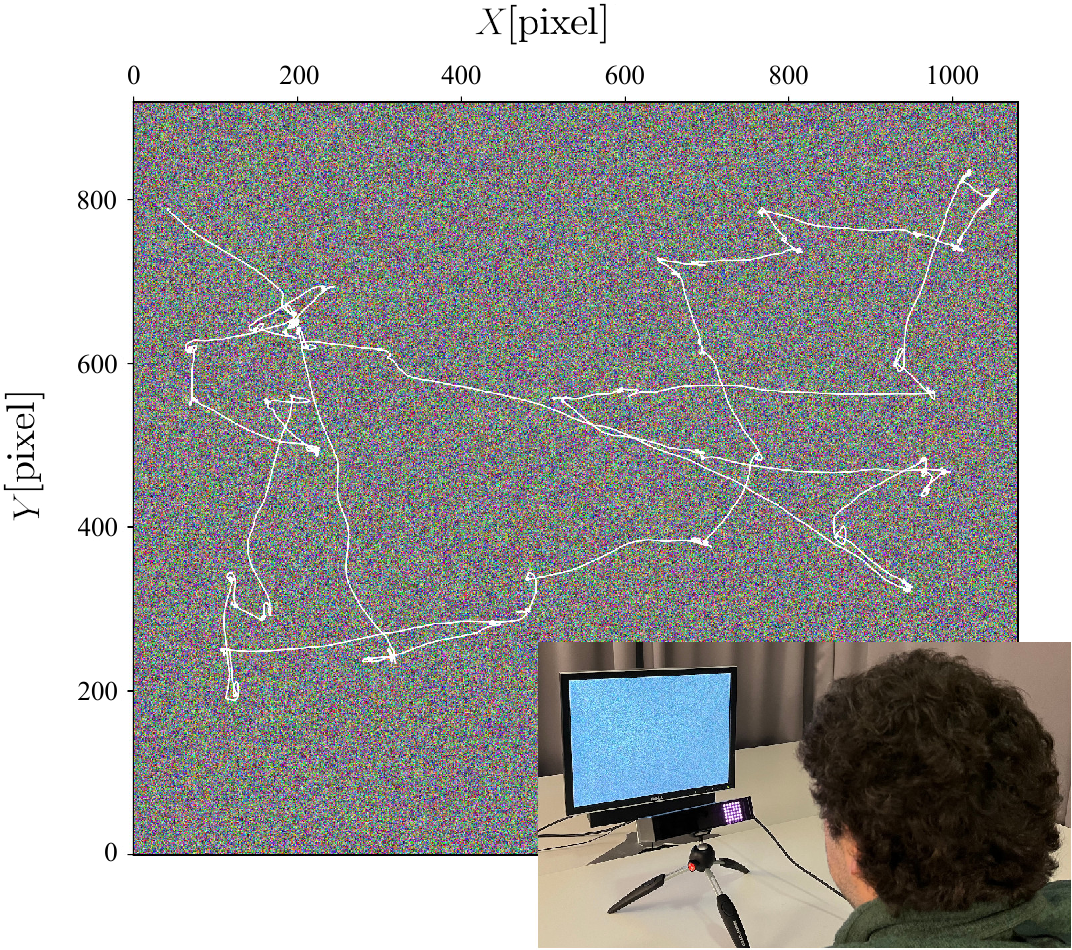}
\end{center}
\caption{Experiment: A subject is presented a random set of $1080 \times 920$ colored pixels for $180$ seconds and asked to find a hidden image. The eye-gaze location is recorded with a rate of $10^3$ frames/second. 
Here a $15$-sec record fragment is shown.}
\label{fig:1}
\end{figure}

Recently, it was demonstrated~\cite{campeau2024intermittent} that some synthetic organisms develop IS strategies rather than Lévy-like ones. This result supports the idea that, while LWs are often adapted as search strategies (e.g., for robots~\cite{fricke2013microbiology,katada2016swarm,duncan2022efficient,krivonosov2016levy}), IS might offer greater efficiency when specific constraints 
are taken into account.

While the two models are essentially different in their statistical properties, distinguishing between these two strategies is not trivial, since the two models produce trajectories that might look very similar~\cite{benichou2011intermittent}. 
The mean squared displacement (MSD) is often used to classify the process~\cite{ariel2015swarming,garg2021efficient,viswanathan1996levy,rhee2011levy}. However, in the realm of limited data, both in terms of quantity and duration, the MSD
alone may not suffice to identify the process. For instance, similar finite-time behavior of the MSD  
can be closely reproduced with both LW and IS. 
Therefore, a stricter classifier is required to effectively address the L\'evy-intermittent dichotomy in an ensemble of finite-length trajectories.

We address this problem in two steps. First, we design an experiment to induce free foraging that allows one to collect a statistically significant number of trajectories. Second, analyzing the collected data we answer the following question: Which model, LW or IS,  best describes \emph{individual} trajectories?

To answer the question, we introduce a moment-based metrics to quantify 
the goodness of fit~\cite{d2017goodness} of a model
and develop a machine learning-assisted algorithm to retrieve model parameters. We find out that, according to the metrics,  most of the trajectories are better described by IS rather than LWs. 

\section{Experimental design} 

The idea to stimulate visual search and analyze recorded gaze trajectories with specific models
has already been 
addressed~\cite{brockmann2000ecology,stephen2009levy,boccignone2004modelling}.
Our experiment differs in two key aspects. Namely, (i) here the search is maximally unconstrained and homogeneous and (ii) we classify not the ensemble but each trajectory individually.

In the experiment, a set of randomly colored pixels is shown to a participant (Fig.~\ref{fig:1}). The participant is asked to find an unspecified hidden visual element during $180$ second, a reasonable duration to balance the level of participant engagement and the need to sample long enough trajectories. 
Our aim was 
to provide conditions for the least constrained visual foraging, confined to a finite area. 
The hardware included an EyeLink Portable Duo eye-tracker~\cite{sr_research_2023}, with a spatial resolution of $0.1$ pixels and a sampling frequency up to $1$ kHz. 

We collected and analyzed data from $120$ subjects, following all legal ethical requirements, and the recorded trajectories were sampled with the time step $\Delta t = 16\text{ ms}$, resulting in $N = 11250$ points per trajectory. 
We label as "not available" (NA), measurements either of subject's blinking or the eye-tracker lost track of the cornea, and discarded records with more than $100$ NAs.

\section{Selecting between L\'evy and Intermittent strategies} 

A planar uniform LW process~\cite{zaburdaev2016superdiffusive} can be described as follows: During time \( \tau_i \), a point-like walker moves with a constant velocity \( v \) in a random direction specified by angle \( \phi \), which is sampled from a uniform distribution on the interval \([0, 2 \pi]\). After completing the ballistic event, a new angle is sampled, and a new duration \( \tau_{i+1} \) is drawn from the PDF
\begin{equation}
\psi(\tau) = \frac{\gamma}{\tau_0} \left(1 + \frac{\tau}{\tau_0}\right)^{-(1+\gamma)}, 
 \label{fPDF}
\end{equation}
with $\gamma\in\ ]0, 2]$ (for $\gamma>2$,  the resulting spread converges to normal diffusion in the asymptotic limit).
Thus, the LW process is parameterized with three positive numbers, $\lbrace v_L, \gamma, \tau_0  \rbrace$. Potentially, $\psi(\tau)$ can be made more flexible, e.g., by modifying its initial part. Our analysis reveals that such modifications do not play a significant role in the classification.

Planar IS~\cite{benichou2005optimal} consists of the two alternating phases mentioned above, the diffusion parameterized by a diffusivity $D$, and the ballistic phase by a constant velocity $v_{\rm B}$.
Together with the two switching rates, the IS is therefore parameterized by the four quantities $\lbrace v_B, D, \lambda_{BD}, \lambda_{DB}\rbrace$.

\begin{figure}[t]
\begin{center}
\includegraphics[width=.5\textwidth]{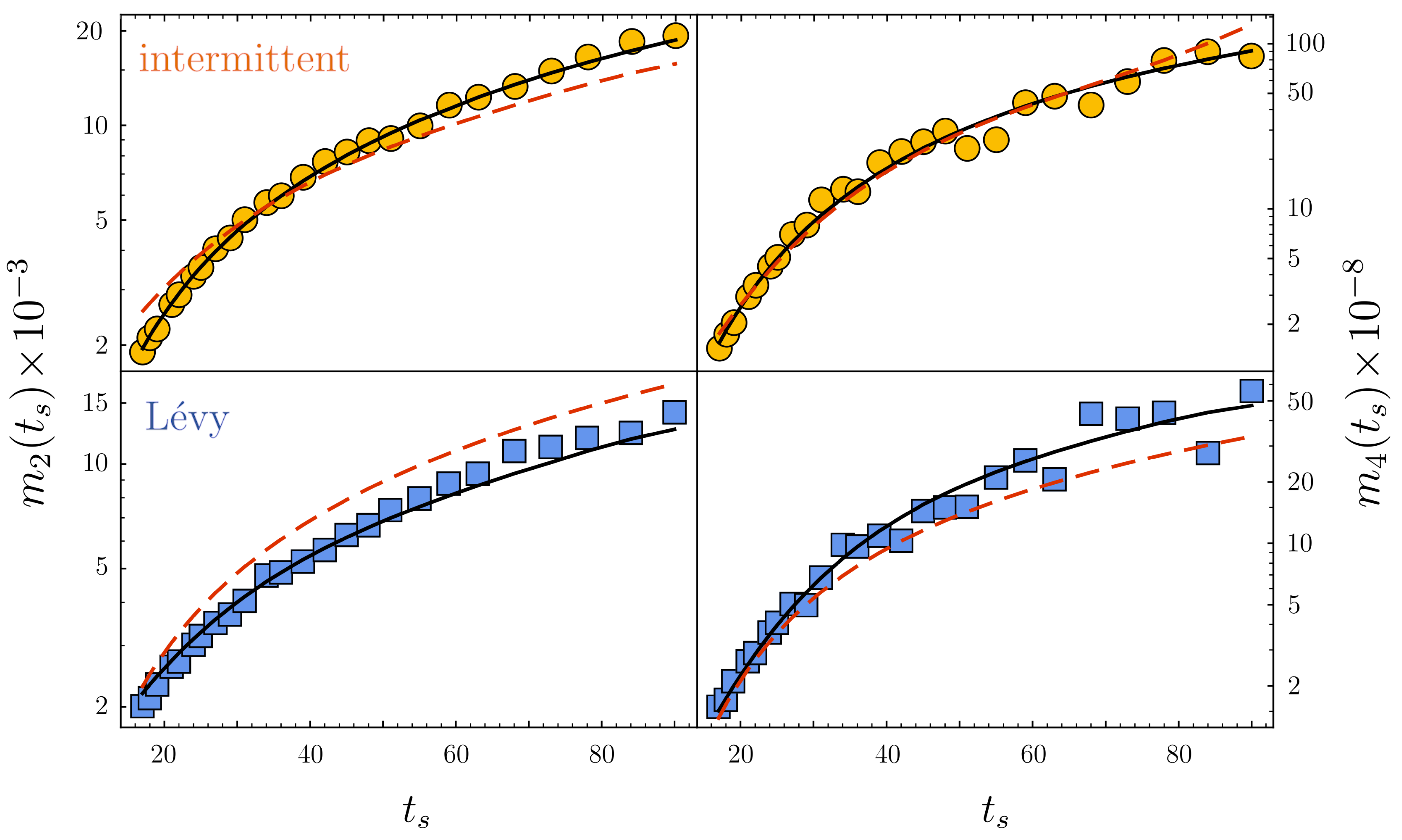}
\end{center}
\caption{
Moments \( m_2(t_s) \) (left), and \( m_4(t_s) \) (right), calculated for the two gaze trajectories with the largest absolute values of \( \Gamma \).
See Fig.~\ref{fig:2}. Moments obtained with the corresponding models and optimal parameter sets (thick solid lines) are shown together with the moments calculated with the complementary models (dashed lines).}
\label{fig3}
\end{figure}

We base our analysis on the second and fourth moments, $m_2(t_s)$ and $m_4(t_s)$ respectively.
To estimate the moments from the recorded time series,  $(X_t,Y_t)$, we first calculate gaze velocities, $v_X(t,t_s) = [X_{t+t_s} - X_{t}]/t_s$ and $v_Y(t,t_s) = [Y_{t+t_s} - Y_{t}]/t_s$.
The moments are then calculated as 
\begin{align}
m_k(t_s) &= \langle \| \Vec{v}(t,t_s) \|^k\rangle_t 
\end{align}
for $k=2,4$ and 
$\| \Vec{v}(t,t_s) \| = \left( v_x^2(t,t_s)+v_y^2(t,t_s)\right)^{1/2}$.
The average is computed by sliding over the data series with a window of the length $t_s$ (see Fig.~\ref{fig3}).


\begin{figure*} [t]
\begin{center}
\includegraphics[width=2.00\columnwidth]{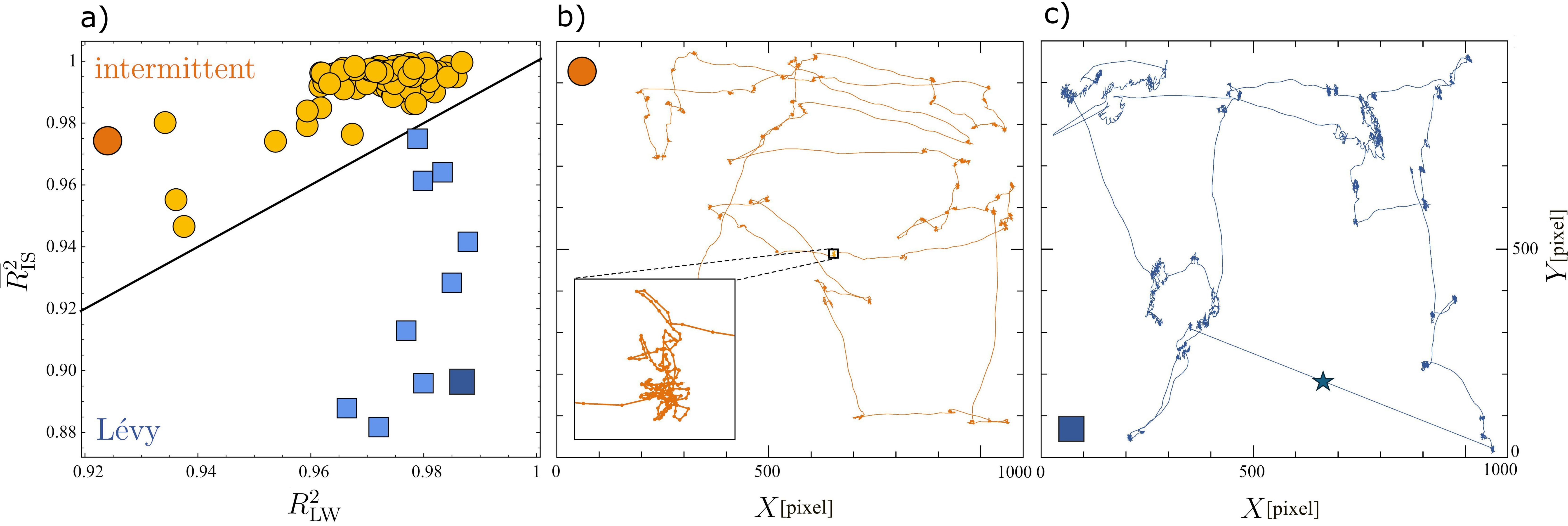}
\end{center}
\caption{Classification of gaze trajectories. (a) Out of the $120$ collected trajectories, $110$ ($\circ$) were classified as produced by intermittent searches and $10$ (\scalebox{0.7}{$\square$}) as produced by L\'evy walks. Classification is based on the adjusted coefficients of determination and a score function 
[cf.~ Eqs.~\eqref{r2} and \eqref{gamma}]. 
Panels (b-c) show 15-second fragments of the trajectories with highest and lowest scores (large circle and square in a), resp.). 
Inset in b): a $90 \, \text{ms}$ fragment.
The star in c) marks a blink event.}
\label{fig:2}
\end{figure*}

The above setup corresponds to the so-called "equilibrated initial conditions"~\cite{klafter1993dynamically}, indicating that the process has already achieved stationarity.
Also, different from LWs, the planar IS allows for full analytical evaluation, and we derived the explicit expressions for the  
moments. 
However, the analytic expression of $m_2(t_s)$ 
alone does not allow to extract all four parameters of the IS
and when analyzing synthetic data, we find that the MSD of $m_2(t_s)$ does not distinguish between the two processes.
This is the reason why we also include $m_4(t_s)$.

To optimize model parameters, we introduce a metric to quantify
the distance between the values of $m_k(t_s)$, 
estimated from the experimental data and 
with each model ${\text{"mod"}} \in \{LW, IS\}$:  
\begin{equation}
    d^{\text{mod}}_{k}(t_s)=\left [ \log
    \left (
    \frac{m^{\text{exp}}_k(t_s)}{m^{\text{mod}}_k(t_s)} 
    \right ) \right ]^2 \, .
\end{equation}
Using the logarithm of the moments 
ensures that the weight is evenly distributed across the values of moments at different times $t_s$. 
The distance is calculated for 25 windows of
size 
$t_s = \lfloor 2^{n/10} \rfloor$ ms 
($n = 41,42,\dots,65$).
The largest value 
\( t_s = 90 \, \text{ms} \), 
is chosen as a compromise between capturing the long-term dynamics 
and minimizing the influence of boundary effects.
We consider $\overline{d}^{\text{mod}}_k$ as the average of $d^{\text{mod}}_k(t_s)$ over $t_s$.

To optimize the parameters of the intermittent search model, we use analytic expressions for $m^{\text{IS}}_k(t_s)$ 
as follows.
We first minimize $\overline{d}^{\text{IS}}_4$, 
using the dual annealing algorithm~\cite{xiang1997generalized} to find an initial guess of all parameters. 
Since 
estimations based on $\overline{d}^{\text{IS}}_4$ minimization alone is highly sensitive to the noise inherent when dealing with finite samples,
we then minimize iteratively $\overline{d}^{\text{IS}}_4$ and $\overline{d}^{\text{IS}}_2$.
Distance $\overline{d}^{\text{IS}}_2$ is used
to update the estiate of $D$ and $\lambda_{DB}$  while keeping $v_B$ and  $\lambda_{BD}$ unchanged, whereas we 
use $\overline{d}^{\text{IS}}_4$ to update the estimate of
$v_B$ and  $\lambda_{BD}$. This Sinkhorn-like~\cite{knight2008sinkhorn}  procedure is repeated iteratively until convergence. 

For the LW model, since the velocity $v$ enters the definitions of the moments in a trivial way, it can be used as a scaling parameter.
The two remaining parameters, $\tau_0$ and $\gamma$, are estimated with a grid search algorithm (see Supplementary Material). 
\begin{figure*}
\begin{center}
\includegraphics[width=0.98\textwidth]{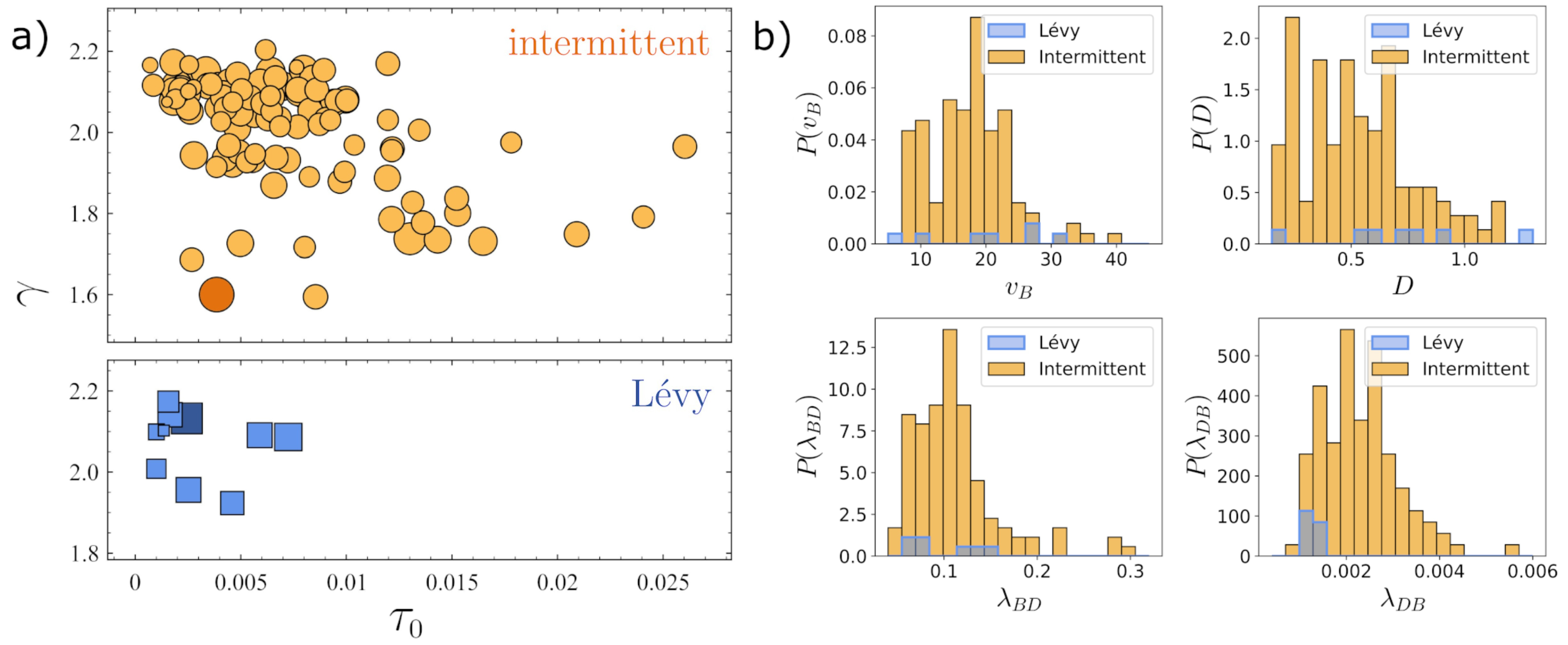}
\end{center}
\caption{\protect
Optimized parameters of the L\'evy walks (a) and intermittent search (b) models for the two gaze trajectories with the largest absolute values of \( \Gamma \).
The area of a circle [a square] is proportional to $1/(1 - {\overline {R}}^{2}_{\text{IS}})$ [$1/(1 - {\overline {R}}^{2}_{\text{LW}})$]. In the case of L\'evy walks, velocity $v$ plays the role of a time scaling parameter.
}
\label{fig:4}
\end{figure*}

To quantify how well each model approximates eye-gaze trajectories, we use the adjusted coefficient of determination~\cite{ezekiel1930methods}, 
\begin{equation}
    {\overline {R}}^{2}_{\text{mod},k}=1-\frac{
\overline{d}^{\text{mod}}_k}
{\sigma^2_{k}} \frac{N-1} {N-p-1} \, ,
\label{r2} 
\end{equation}
where $N=25$, $\sigma^2_{k}$ represents the variance of $\log[m^{\text{exp}}_k(t_s)]$ over the $t_s$-values, 
and $p=3,4$ is the number of free parameters of LW and IS model respectively. 
Finally, 
we define  the score function,
\begin{equation}
     \Gamma = \overline {R}^{2}_{\text{IS}} - \overline {R}^{2}_{\text{LW}}  ,
     \label{gamma}   
 \end{equation}
with 
$\overline {R}^{2}_{\text{mod}} = (\overline {R}^{2}_{\text{mod},2} + \overline {R}^{2}_{\text{mod},4})/2$.
We choose model IS if $\Gamma > 0$ and LW other wise. See Fig.~\ref{fig:2}.

We have benchmarked the algorithm with $10^3$ synthetic trajectories of each model, for a wide range parameter. The trajectories were sampled with the same frequency and with the same number of record points as in the experiment. The algorithm was able to correctly classify $996$ (LWs) and $980$ (IS) trajectories, recovering also the model parameters.
Figure \ref{fig3} illustrates the mismatch between both models for synthetic data. 


\section{Discussion and outlook} 

Figure \ref{fig:2} presents the results of classification of $120$ trajectories collected in the experiment: 
$110$ trajectiries  are classified as IS and only $10$ as LWs. 
This finding corroborates the conventional (qualitative) description of eye-gaze dynamics as processes consisting of two alternating phases, ballistic-like saccades and diffuse-like fixations~\cite{liversedge2011oxford, yarbus2013eye, lencastreDataScience}.
However, we notice that both IS and LW present good fitness, showing values of $\overline{R}^2_{IS}$ and $\overline{R}^2_{LW}$ close to $1$.
The values of their differences $\Gamma$, are comparable with the values obtained with the synthetic data. 

It is noteworthy that, from the LW  perspective, nearly all trajectories [except for the four leftmost in Fig.~\ref{fig:2}(a)] have $\overline{R}^2_{LW}$ values within the same range. It is the $\overline{R}^2_{LW}$ coefficient that reveals that most of the trajectories are actually much better approximated by the intermittent search model. 

Finally, Figure~\ref{fig:4} presents the values of the parameters of each model for every individual trajectory, indicating again with different symbols which model fits best. From the LW  perspective,
the trajectories are characterized by values of the exponent $\gamma\sim 2$ (Fig.~\ref{fig:4}a), i.e., all classified LWs lay between diffusive and superdiffusive regimes~\cite{zaburdaev2015levy}, which should be a result of the specific setup of our experiment. Under different conditions, one could expect gaze trajectories with other exponent values when being modeled as LWs~\cite{levernier2020inverse}.

From the IS perspective, trajectories are characterized by distributions of the four parameters, $\lbrace D, v_B, \lambda_{BD}, \lambda_{DB}\rbrace$ (Fig.~\ref{fig:4}b).  The duration of saccades is known to lay in the range between $8$ and $60 \, \text{ms}$~\cite{lencastreDataScience, podladchikova2009temporal, galley2015fixation, devillez2020bimodality}, which is in good agreement with estimated values of $1/\lambda_{BD}$, shown in Fig.~\ref{fig:4}.
In the same way, the distribution of values of $1/\lambda_{DB}$ reproduces fixation durations, known to lay in the range between $90$ and $800 \, \text{ms}$. 


We have analyzed eye-gaze trajectories collected in an experiment stimulating free search for visual information. Using an algorithm designed to quantify the ability of IS and LW models to approximate a given trajectory, based on the second and fourth moments of eye-gaze displacements, we found that the IS model scores significantly higher than the LW in most cases.
Our results depend on the classification setup  is based on a comparison of the performance of each model to fit both moments.

We use the models in their most basic formulations. Both can be further modified, e.g., by adding a stochastic component to the ballistic phases~~\cite{zaburdaev2011perturbation}, introducing a vertical/horizontal non-uniformity, substituting
the uniform planar LWs by intermittent L\'evy walks~\cite{guinard2021intermittent}, or, in an extreme case, considering composite random walks~\cite{morales2004extracting,reynolds2008many}.
Moreover, the fact that the search is restricted to a bounded domain (the computer screen) is not taken into account, even though this might influence statistical characteristics of the process. 
While these amendments might bring the models close to the realm of the experiment, we do not consider them here because our objective is to formulate a \textit{well-posed} dichotomous problem. 

Finally, we also emphasize the idea of score-based model selection, which we believe enables us to transform the ill-posed problem of identifying the 'best model' into a well-defined, implementable approach that works on the level of individual trajectories.
Several steps in this direction have already been made. For instance, likelihood-based methods have been proposed to distinguish between different diffusion processes~\cite{auger2015differentiating}. Additionally, machine-learning-based approaches, as highlighted in the recent AnDi Challenge~\cite{munoz2021objective}, may offer powerful tools for score-based model selection at the single-trajectory level.
Score-based classification of individual trajectories will allow replacing the task of 'macroscopic' en masse classification of foraging patterns~\cite{ariel2015swarming,sims2008scaling,humphries2010environmental,brown2007levy,raichlen2014evidence,garg2021efficient} with an analysis of the distributions of scores and optimized parameters, similar to the results presented in Figs.~\ref{fig:2} and \ref{fig:4}.  This microscopic approach could offer deeper insights into the statistical features of the patterns and thus would help to understand the underlying foraging dynamics better.

The authors thank Research Council of Norway, under the project “VirtualEye" (Ref.~335940-FORSKER22) and Oslo Metropolitan University, under the initiative {\it NordSTAR}, for partial support.



%
\begin{widetext}

\makeatletter \renewcommand{\fnum@figure}{{\bf{\figurename~S\thefigure}}}
\setcounter{figure}{0}
\setcounter{equation}{0}
\renewcommand{\theequation}{S\arabic{equation}}
\section{Supplementary Material}
\setcounter{section}{0}
\setcounter{table}{0}
\renewcommand{\thetable}{S\arabic{table}}

\subsection{Analytical expressions for the moments of an intermittent process}

\subsection{Preliminary assumptions and derivations}

Consider the continuous-time random walk process characterized by random movements in a two-dimensional space, random waiting times between these movements, and can be in two phases, either ballistical $(B)$ or diffusive $(D)$
Further, for the convenience of deriving equations, we will denote phases $B$ and $D$ as $``+"$ or $``-"$, respectively.  We can consider this process as a particle performing a random wandering.

Let $P(\mathbf{r},t)$ be the probability density function (pdf) of finding the particle at a point  $\mathbf{r}$ measuring at the moment of time $t$.
We can write $P(\mathbf{r},t)=P_+(\mathbf{r},t)+P_-(\mathbf{r},t)$,
where $P_{\pm}(\mathbf{r},t)$ is a density that at the moment $t$ the particle is located at a coordinate $\mathbf{r}$ being in phases $``+"$ or $``-"$.
In addition, we introduce the pdf $\nu(\mathbf{r},t)$ that the particle is in a space-time coordinate $(\mathbf{r},t)$.
It is important to note that $P(\mathbf{r},t)$  is the pdf concerning a coordinate $\mathbf{r}$ only (time $t$ plays a role of the parameter), but $\nu(\mathbf{r},t)$ is the pdf over both variables $\mathbf{r}$ and $t$.
Analogously, we can decompose $\nu(\mathbf{r},t)$ as $\nu(\mathbf{r},t)=\nu_{+}(\mathbf{r},t)+\nu_{-}(\mathbf{r},t)$.
The density $\nu_{\pm}(\mathbf{r},t)$ can be interpreted as a flux describing the fraction of particles that just have arrived  (per unit of time) at position $\mathbf{r}$ at time $t$ and thereby making switching from the phase $``\mp"$  to phase  $``\pm"$.
We also assume that at the beginning of the wandering, the particle is located at the origin while staying in the phase $``\pm"$  with the probability $c_{\pm}$ ($c_+ + c_-=1$). In the ensemble interpretation, $c_{\pm}$ is the fraction of the particles initially in the phase $``\pm"$.

Next, we denote by $\mathrm{d}\nu_{\pm} (\mathbf{r},t|\mathbf{r}^{\prime},t^{\prime})$ the particular flux per unit of time from phase  $``\mp"$  and from fixed point  $(\mathbf{r}^{\prime},t^{\prime})$ to phase $``\pm"$ and point $(\mathbf{r},t)$. The total flux
$\nu_{\pm}(\mathbf{r},t)$ is given by the integral of the particular flux over $\mathbf{r}^{\prime}$ and $t^{\prime}$, namely,
\begin{equation}
\label{nu}
    \nu_{\pm}(\mathbf{r},t) =c_{\mp}w_{\mp}(\mathbf{r},t)\psi_{\mp}(t) + \int\mathrm{d}\nu_{\pm}(\mathbf{r},t|\mathbf{r}^{\prime},t^{\prime}),
\end{equation}
where the integration volume is defined by the condition $\mathbf{r}^{\prime} \in \mathbb{R}^2$ and $0 \leq t^{\prime}\leq t$.
The first term captures the scenario where the particle has changed its phase only once (from $``\mp"$ to  $``\pm"$). In this case, time which the particle spent in the current phase is equal to \(t\), and at the very end, the particle switches to the alternate phase.

Let us take it into consideration the function $\psi_{\mp}(t^{\prime})$ that is the pdf of being in a phase $``\mp"$ during time $t^{\prime}$.
Then, probability that a phase transition, from  $``\mp"$ to  $``\pm"$, happened  in time interval $(t^{\prime},t^{\prime}+\mathrm{d}t^{\prime})$ is equal to $\psi_{\mp}(t^{\prime})\mathrm{d}t^{\prime}$.
Similarly, the probability that at the moment $t^{\prime}$ the particle movement in the phase $``\mp"$ has a length within the interval $(\mathbf{r}^{\prime},\mathbf{r}^{\prime}+\mathrm{d}\mathbf{r}^{\prime})$ is written as $w_{\mp}(\mathbf{r}^{\prime},t^{\prime})\mathrm{d}\mathbf{r}^{\prime}$, where  $w_{\mp}(\mathbf{r}^{\prime},t^{\prime})$ are the corresponding pdf (with respect to the variable $\mathbf{r}^{\prime}$).
Here and below, of course, we assume  $\mathrm{d}\mathbf{r}^{\prime} \rightarrow 0$   and $\mathrm{d}t^{\prime} \rightarrow 0$.

As a result, the particular flux $\mathrm{d}\nu_{\pm}(\mathbf{r},t|\mathbf{r}^{\prime},t^{\prime})$ is represented as a product of probabilities  $\psi_{\mp}(t^{\prime})\mathrm{d}t^{\prime}$ and $w_{\mp}(\mathbf{r}^{\prime},t^{\prime})\mathrm{d}\mathbf{r}^{\prime}$ multiplied by total flux $\nu_{\mp}(\mathbf{r}-\mathbf{r}^{\prime},t-t^{\prime})$,  namely
\begin{equation}
\label{nu_2}
   \mathrm{d}\nu_{\pm}(\mathbf{r},t|\mathbf{r}^{\prime},t^{\prime})=
   \nu_{\mp}(\mathbf{r}-\mathbf{r}^{\prime},t-t^{\prime})
   w_{\mp}(\mathbf{r}^{\prime},t^{\prime})\mathrm{d}\mathbf{r}^{\prime}
   \psi_{\mp}(t^{\prime})\mathrm{d}t^{\prime}.
\end{equation}
The meaning of this formula is that the conditional probability density to move from the point $(\mathbf{r}^{\prime},t^{\prime})$ to the point $(\mathbf{r},t)$ and to switch out of phase $``\mp"$ to phase $``\pm"$ at the end of the movement is equal to the product of the probability density of arriving at the point $(\mathbf{r}-\mathbf{r}^{\prime}, t-t^{\prime})$ in the phase $``\mp"$, and the joint the probability that the remaining time in the phase $``\mp"$  is equal to $t^{\prime}$ and the displacement has a length of $\mathbf{r}^{\prime}$, after which the particle switch to the state $``\pm"$. Thus, after integration of Eq.~\eqref{nu_2}, we get
\begin{equation}
\label{2D:nu1}
    \nu_{\pm}(\mathbf{r},t)=c_{\mp}\psi_{\mp}(t)w_{\mp}(\mathbf{r},t) +
    \int_{0}^{t}\mathrm{d}t^{\prime}\psi_{\mp}(t^{\prime})
    \int_{\mathbb{R}^2}\mathrm{d}\mathbf{r}^{\prime}w_{\mp}(\mathbf{r}^{\prime},t^{\prime})
   \nu_{\mp}(\mathbf{r}-\mathbf{r}^{\prime},t-t^{\prime}).
\end{equation}

Based on a similar idea, one can determine the probability density $P_{\pm}(\mathbf{r},t)$, which represents the pdf that a particle is located at point $\mathbf{r}$ and in the phase $``\pm"$ at time $t$.
This density is calculated as an integral over all possible realizations of the process under consideration. The integral involves the flux  $\nu_{\pm}(\mathbf{r}-\mathbf{r}^{\prime},t-t^{\prime})\mathrm{d}t^{\prime}$ which describes  the fraction of particles positioned at the coordinate $\mathbf{r}-\mathbf{r}^{\prime}$ at the moment of time  from the interval $(t-t^{\prime},t-t^{\prime}+\mathrm{d}t^{\prime})$
multiplied by the probability that the time spent in the current phase $``\pm"$ exceeds $t^{\prime}$, denoted as $\Psi_{\pm}(t^{\prime})=\int_{t^{\prime}}^{\infty}d\tau\psi_{\pm}(\tau)$, and by $w_{\pm}(\mathbf{r}^{\prime},t^{\prime})\mathrm{d}\mathbf{r}^{\prime}$ which is a conditional probability that displacement over time $t^{\prime}$ along the spatial coordinate falls within the interval $(\mathbf{r}^{\prime},\mathbf{r}^{\prime}+\mathrm{d}\mathbf{r}^{\prime})$. Therefore, the equation that gives the pdfs $P_{\pm}(\mathbf{r},t)$ takes a form
\begin{equation}
\label{2D:P1}
    P_{\pm}(\mathbf{r},t)=c_{\pm}\Psi_{\pm}(t)w_{\pm}(\mathbf{r},t)
    +
    \int_{0}^{t}\mathrm{d}t^{\prime}\Psi_{\pm}(t^{\prime})
    \int_{\mathbb{R}^2}\mathrm{d}\mathbf{r}^{\prime}w_{\pm}(\mathbf{r}^{\prime},t^{\prime})
   \nu_{\pm}(\mathbf{r}-\mathbf{r}^{\prime},t-t^{\prime}).
\end{equation}
The first term in the latter expression indicates the situation when the particles remained in the initial phase.

For convenience, let us denote
\begin{equation}
\label{2D:q_Q}
    q_{\pm}(\mathbf{r},t)=\psi_{\pm}(t)w_{\pm}(\mathbf{r},t), \quad
    Q_{\pm}(\mathbf{r},t)=\Psi_{\pm}(t)w_{\pm}(\mathbf{r},t).
\end{equation}
Then, equations \eqref{2D:nu1} and \eqref{2D:P1} can be rewritten as
\begin{align}
    \label{2D:nu2}
    \nu_{\pm}(\mathbf{r},t)&=c_{\mp}q_{\mp}(\mathbf{r},t) + \int_{0}^{t}\mathrm{d}t^{\prime}  \int_{\mathbb{R}^2}\mathrm{d}\mathbf{r}^{\prime}q_{\mp}(\mathbf{r}^{\prime},t^{\prime})
   \nu_{\mp}(\mathbf{r}-\mathbf{r}^{\prime},t-t^{\prime}),
    \\
   P_{\pm}(\mathbf{r},t)&=c_{\pm}\Psi_{\pm}(t)w_{\pm}(\mathbf{r},t)+ \int_{0}^{t}\mathrm{d}t^{\prime}
    \int_{\mathbb{R}^2}\mathrm{d}\mathbf{r}^{\prime}Q_{\pm}(\mathbf{r}^{\prime},t^{\prime})
   \nu_{\pm}(\mathbf{r}-\mathbf{r}^{\prime},t-t^{\prime}).
   \label{2D:P2}
\end{align}
As a consequence, in the Fourier-Laplace space we obtain
\begin{align}
    \nu_{\pm}(\mathbf{k},s) &=q_{\mp}(\mathbf{k},s)[c_{\mp} + \nu_{\mp}(\mathbf{k},s)],
    \label{2D:nu_ks}\\
    P_{\pm}(\mathbf{k},s) &=Q_{\pm}(\mathbf{k},s)[c_{\pm}+\nu_{\pm}(\mathbf{k},s)], \label{2D:P_ks}
\end{align}
where argument $\mathbf{k}$ indicates the Fourier transform of the respective function with argument $\mathbf{r}$, and argument $s$ indicates the Laplace transform of the respective function with argument $t$, i.e.,
\begin{align}
    f(\mathbf{k})&=\mathcal{F}_{\mathbf{r}\to \mathbf{k}}\left\{f(\mathbf{r})\right\}=\int_{\mathbb{R}^2} e^{i \mathbf{k}\cdot\mathbf{r}}f(\mathbf{r})\mathrm{d}\mathbf{r},
    \nonumber\\
    g(s)&=\mathcal{L}_{t\to s}\left\{g(t)\right\}=\int_{0}^{\infty} e^{-s t}g(t)\mathrm{d}t.
    \nonumber
\end{align}

Solving the system of equations \eqref{2D:nu_ks} and \eqref{2D:P_ks} with respect to $\nu_{\pm}(\mathbf{r},t)$, we find
\begin{equation}
\label{2D:nu_ks1}
    \nu_{\pm}(\mathbf{k},s) =q_{\mp}(\mathbf{k},s)\frac{c_{\mp} +c_{\pm}q_{\pm}(\mathbf{k},s)}{1-q_{+}(\mathbf{k},s)q_-(\mathbf{k},s)}.
\end{equation}
Next, taking into account Eqs.~\eqref{2D:nu_ks1} and \eqref{2D:P_ks}, we have
\begin{equation}
\label{2D:P_ks1}
    P_{\pm}(\mathbf{k},s) =Q_{\pm}(\mathbf{k},s)\frac{c_{\pm}+ c_{\mp}q_{\mp}(\mathbf{k},s)}{1-q_{+}(\mathbf{k},s)q_-(\mathbf{k},s)}.
\end{equation}
Therefore, we have the required propagator in the Fourier-Laplace space
\begin{align}
\label{2D:P_ks2}
    P(\mathbf{k},s)&=
    \frac{Q_{+}(\mathbf{k},s)[c_{+}+c_{-}q_{-}(\mathbf{k},s)] + Q_{-}(\mathbf{k},s)[c_{-}+c_{+}q_{+}(\mathbf{k},s)]}{1-q_{+}(\mathbf{k},s)q_-(\mathbf{k},s)}.
\end{align}

Note that we have provided the derivation of the propagator \eqref{2D:P_ks2} for our process in the case of a $2$--dimensional random walk. However, this derivation is obviously correct for any dimension of space.

To derive the final expression for the propagator, which will allow us to calculate the second and fourth moments of the process, we first write down
the probability density of the duration of each of the phases, ballistic ($B$) and diffusive ($D$), specifically
\begin{equation}
\label{2D:psi}
    \psi_B(t)=\lambda_B e^{-\lambda_B t}, \quad
    \psi_D(t)=\lambda_D e^{-\lambda_D t} .
\end{equation}
In the main text of the paper we denote the transition rates from state $B$ to $D$ and from state $D$ to $B$ as $\lambda_{BD}$ and $\lambda_{DB}$, respectively. Here, for compactness of the notation, we denote $\lambda_{BD}$ as $\lambda_{B}$ and $\lambda_{DB}$ as $\lambda_{D}$.
Next, for our isotropic case, the conditional probabilities that the particle being in each phase ($B$ or $D$) has the position $\mathbf{r}$ at time $t$ are given by
\begin{equation}
\label{2D:w_D-w_B}
    w_B(\mathbf{r},t)=\frac{\delta(r-v_Bt)}{2\pi r} ,
    \quad
    w_D(\mathbf{r},t)=\frac{1}{4\pi D t}\exp\!\left(-\frac{r^2}{4D t}\right) ,
\end{equation}
where $r=|\mathbf{r}|$.

We now can write expressions from Eqs.~\eqref{2D:q_Q}, replacing the indices $``\pm"$ with $``B"$ and $``D"$, and substituting the probability densities written in Eqs.~\eqref{2D:psi} and \eqref{2D:w_D-w_B}, as
\begin{align}
  \label{2D:q-Q_BD}
    q_B(\mathbf{k},s)&=\frac{\lambda_B}{\sqrt{(s+\lambda_B)^2+v_B^2 k^2}}, \quad\,\,
    q_D(\mathbf{k},s)=\frac{\lambda_D}{s+\lambda_D+Dk^2} , \nonumber\\
    Q_B(\mathbf{k},s)&=\frac{1}{\sqrt{(s+\lambda_B)^2+v_B^2 k^2}} , \quad
    Q_D(\mathbf{k},s)=\frac{1}{s+\lambda_D+Dk^2}  ,
\end{align}
where $k=|\mathbf{k}|$.
Finally, using these formulas and Eq.~\eqref{2D:P_ks2}, we obtain  the propagator
\begin{align}
  \label{2D:P_ks3}
    P(\mathbf{k},s)
    &=
    \frac{c_B(s+\lambda_D+Dk^2)+c_D\sqrt{(s+\lambda_B)^2+v_B^2 k^2}+c_B\lambda_B+c_D\lambda_D}
    {(s+\lambda_D+Dk^2)\sqrt{(s+\lambda_B)^2+v_B^2 k^2}-\lambda_B \lambda_D}.
\end{align}

\subsubsection{Derivation of the second moment}

The second moment of the intermittent process can be written as
\begin{eqnarray}
   \langle \mathbf{r}^2(t)\rangle = \int_{\mathbb{R}^2} r^2 P(\mathbf{r},t)\mathrm{d}\mathbf{r}
   =\mathcal{L}^{-1}_{t\to s}\left\{-\left(\frac{\partial^2}{\partial k_x^2}+\frac{\partial^2}{\partial k_y^2}\right)P(\mathbf{k},s)\big|_{\mathbf{k}=0}\right\} .
\label{2D:x2}
\end{eqnarray}
Using expressions \eqref{2D:P_ks3} and \eqref{2D:x2}, along with the application of the inverse Laplace transform, we can derive the second moment  as
\begin{equation}
\label{2D:x22}
   \langle \mathbf{r}^2(t)\rangle=\mathcal{C}^{(2)}_1t+\mathcal{C}^{(2)}_2+\mathcal{C}^{(2)}_3e^{-\lambda_B t}
   +\mathcal{C}^{(2)}_4e^{-(\lambda_B+\lambda_D) t},
\end{equation}
where constants
\begin{align}
\label{2D:C}
   \mathcal{C}^{(2)}_1&=2\frac{2D\lambda_B^2+v_B^2\lambda_D}{\lambda_B(\lambda_B+\lambda_D)}  ,
   \nonumber\\
   \mathcal{C}^{(2)}_2&=\frac{2}{\lambda_B^2(\lambda_B+\lambda_D)^2} \left\{
     2D \lambda_B^2(\lambda_D c_D-\lambda_B c_B)+v_B^2[\lambda_B^2 c_B-\lambda_B \lambda_D (1+c_D)-\lambda_D^2] \right\} ,
    \nonumber\\
   \mathcal{C}^{(2)}_3&=2\frac{v_B^2(\lambda_D-\lambda_B c_B)}{\lambda_B^2\lambda_D}  ,
   \nonumber\\
   \mathcal{C}^{(2)}_4&=2\frac{(2D \lambda_D+v_B^2)(\lambda_B c_B-\lambda_D c_D)}
   {\lambda_D (\lambda_B+\lambda_D)^2}.\nonumber
\end{align}
Let us remind that $c_B$ and $c_D$ are the initial probabilities to start the process in the ballistic or diffusive phase, respectively.
For equilibrated starting conditions, we have
\begin{equation}
\label{2D:equilib}
    c_B = \frac{\lambda_D}{\lambda_D+\lambda_B}, \quad
    c_D = \frac{\lambda_B}{\lambda_D+\lambda_B} .
\end{equation}
In this case, the second moment simplifies to
\begin{equation}
   \langle \mathbf{r}^2(t)\rangle = C_1^{(2)} t-C_{2}^{(2)}(1-e^{-\beta\alpha t}) \,
\label{eq:secondmomentinterm}
\end{equation}
with the values of the constants
\begin{eqnarray}
    C_1^{(2)} &=& 2 \frac{1-\alpha}{\alpha} D_I
      + 4\alpha D  ,\nonumber\\
      &   & \cr
    C_2^{(2)} &=& 2 \frac{1}{\beta\alpha} \frac{1-\alpha}{\alpha}D_I.\nonumber
\end{eqnarray}
Here, we introduce the following parameters
\begin{equation}
   \beta = \lambda_{B}+  \lambda_{D}, \quad \alpha = \frac{\lambda_{B}}{\beta}, \quad D_I=\frac{v_B^2}{\beta}.
\label{eq:parameters}
\end{equation}

\subsubsection{Derivation of the fourth moment}
\label{appendfourth}

The fourth-order moment can be calculated similarly to the second-order moment.  At first, we express it as
\begin{eqnarray}
   \langle \mathbf{r}^4(t)\rangle &=& \int_{\mathbb{R}^2} r^4 P(\mathbf{r},t)\mathrm{d}\mathbf{r}
   =\mathcal{L}^{-1}_{t\to s}\left\{\left(\frac{\partial^4}{\partial k_x^4}+\frac{\partial^4}{\partial k_y^4}\right)P(\mathbf{k},s)\big|_{\mathbf{k}=0}\right\} .
\label{2D:x4}
\end{eqnarray}
Then, using the expression for \(P(\mathbf{k}, s)\) found in Eq.~\eqref{2D:P_ks3}, after applying the inverse Laplace transform, we obtain
\begin{equation}
\label{2D:4moment}
   \langle \mathbf{r}^4(t)\rangle
   =\mathcal{C}^{(4)}_1t^2+\mathcal{C}^{(4)}_2 t+\mathcal{C}^{(4)}_3+\mathcal{C}^{(4)}_4t^2e^{-\lambda_B t}
   +\mathcal{C}^{(4)}_5te^{-\lambda_B t}+\mathcal{C}^{(4)}_6te^{-(\lambda_B+\lambda_D) t}
   +\mathcal{C}^{(4)}_7e^{-\lambda_B t}+\mathcal{C}^{(4)}_8e^{-(\lambda_B+\lambda_D) t}
\end{equation}
with
\begin{eqnarray}
    \mathcal{C}^{(4)}_1  &=& \frac{8}{\lambda_B^2 (\lambda_B + \lambda_D)^2}
\left[2 D\lambda_B^2 + v_B^2\lambda_D\right]^2   ,
\nonumber\\
& & \cr
\mathcal{C}^{(4)}_2  &=& \frac{8}{\lambda_B^3 (\lambda_B + \lambda_D)^3} \big\{
    -8c_B D^2 \lambda_B^5
    +4\big[2(1+c_D) D^2\lambda_D + c_B v_B^2 D\big]\lambda_B^4-20v_B^2 D\lambda_D\lambda_B^3
    \cr
    & & \cr
    & & + \big[v_B^2(2c_B+3)-4D(c_B +1)\lambda_D\big]v_B^2\lambda_D\lambda_B^2
    -2 v_B^4(1+c_D)\lambda_D^2\lambda_B
    -3\lambda_D^3 v_B^4
\big\}  ,
\nonumber\\
& & \cr
\mathcal{C}^{(4)}_3  &=& \frac{8}{\lambda_B^4 (\lambda_B + \lambda_D)^4} \big\{
    8c_B D^2\lambda_B^6
    -(16D^2\lambda_D+12 D c_B v_B^2)\lambda_B^5
    +\big[3c_Bv_B^4 + 4v_B^2(8+c_D)D\lambda_D + 8c_DD^2\lambda_D^2\big]\lambda_B^4 \cr
    & & \cr
    & &+v_B^2\big[(12c_B+16)D\lambda_D - v_B^2(12+c_B)\big]\lambda_D \lambda_B^3
    +\big[4(1+c_B)D\lambda_D - (3+7c_B)v_B^2\big]v_B^2\lambda_D^2\lambda_B^2
    \cr
    & & \cr
    & &
    +3v_B^4(1+c_D)\lambda_D^3\lambda_B + 3v_B^4 \lambda_D^4
\big\}  ,
\nonumber\\
& & \cr
\mathcal{C}^{(4)}_4 &=&\frac{4}{\lambda_B^2 \lambda_D}
    v_B^4(-c_B \lambda_B +\lambda_D)  ,
\nonumber\\
& & \cr
\mathcal{C}^{(4)}_5  &=& \frac{8 v_B^2 }{\lambda_B^2 \lambda_D^2}
\big[-c_B v_B^2 \lambda_B + (1+c_B)v_B^2\lambda_D\big]  ,
\nonumber\\
& & \cr
\mathcal{C}^{(4)}_6  &=& \frac{16}{\lambda_D^2 (\lambda_B + \lambda_D)^3} [
    (-2c_B D v_B^2 \lambda_D -c_B v_B^4)\lambda_B^2
    +(c_D v_B^4 + 4c_B D^2 \lambda_D^2 +2v_B^2 D\lambda_D\big)\lambda_D\lambda_B \cr
    & & \cr
    & &
    - 2 c_D D (v_B^2 + 2 D \lambda_D)\lambda_D^3 ]  ,
\nonumber\\
& & \cr
\mathcal{C}^{(4)}_7  &=& \frac{8}{\lambda_B^4 \lambda_D^3} \big\{
    -(3 v_B^2  + 4D^2\lambda_D) c_B v_B^2\lambda_B^3
    +[(3+5c_B)v_B^2+(1+c_B)4D\lambda_D]v_B^2\lambda_D\lambda_B^2 \cr
    & & \cr
    & &
    -3(2+c_B)v_B^4\lambda_D^2\lambda_B + 3v_B^4 \lambda_D^3
\big\}  ,
\nonumber\\     & & \cr
\mathcal{C}^{(4)}_8 &=&\frac{8}{\lambda_D^3  (\lambda_B+\lambda_D)^4} \big\{
    (3v_B^2+4D\lambda_D)c_B v_B^2 \lambda_B^3
    +[(2c_B-c_D)4v_B^2D\lambda_D+(4c_B-3c_D)v_B^4+8c_B D^2\lambda_D^2]\lambda_D\lambda_B^2
    \cr
    & & \cr
    & &-[4v_B^2(4+c_B)D\lambda_D+16D^2 \lambda_D^2+(5+c_D)v_B^4]\lambda_D^2\lambda_B
    +c_D[8D^2 \lambda_D^2+3v_B^4+12 Dv_B^2\lambda_D]\lambda_D^3
   \big\}.
   \nonumber
\end{eqnarray}
Once again, if we assume equilibrated starting conditions, Eq.~\eqref{2D:equilib}, the equation above reduces to
\begin{eqnarray}
 \langle \mathbf{r}^4(t)\rangle
    &=& C_1^{(4)}t^2+C_2^{(4)} t-C_3^{(4)} - C_4^{(4)}e^{-\beta t}  +  \left ( C_5^{(4)}t^2 + C_6^{(4)}t -C_7^{(4)} \right ) e^{-\beta\alpha t} \label{eq:fourthmomentinterm}
\end{eqnarray}
with
\begin{eqnarray}
    C_1^{(4)}  &=&  8\alpha^2 \left( 2D+ 
    \frac{1-\alpha}{\alpha^2} D_I    \right)^2 , \nonumber\\
    & & \cr
    C_2^{(4)}  &=& 8\frac{1}{\beta^2\alpha^2}
                    \frac{1-\alpha}{\alpha}
                 \left[ \left( D-\frac{\alpha+1}{2\alpha^2} D_I \right)^2 - \frac{5}{8\alpha^4} D_I^2 \right] , \nonumber\\
    & & \cr
    C_3^{(4)}  &=& 4\frac{1}{\beta^4\alpha^4} (1-\alpha) \left[ \left ( 2D-\frac{\alpha^2+\alpha+1}{\alpha^3} D_I \right )^2 \right .  \left . - \frac{3\alpha^3-3\alpha^2-\tfrac{\alpha}{2}-1}{\alpha^6} \right]  , \nonumber\\
    & & \cr
    C_4^{(4)}  &=& 32\beta\alpha(1-\alpha) \left ( 2D+\frac{1}{1-\alpha} D_I    \right )^2  ,  \nonumber\\
    & & \cr
    C_5^{(4)}  &=&  4\frac{(1-\alpha)}{\alpha^2} D_I^2  ,  \nonumber\\
    & & \cr
    C_6^{(4)}  &=&  \frac{16}{\alpha}\frac{1}{\beta\alpha} D_I^2   , \nonumber\\
    & & \cr
    C_7^{(4)}  &=&  \frac{8}{\beta^2\alpha^2} D_I \left ( 8D+\frac{11\alpha^2-12\alpha+3}{\alpha^2 (1-\alpha)} D_I    \right )^2  . \nonumber
\end{eqnarray}

\subsection{Choosing the best model and estimating its parameters}

\subsubsection{The L\'evy walk model}


In the L\'evy walk model, the $k$-moment of the displacements, $m_k^{LW(v_L,\gamma,\tau_0)}(t_s)$, scales with the velocity magnitude  as follows:
\begin{align}
m_k^{LW(v_L,\gamma,\tau_0)}(t_s) & =  v_L^k m_k^{LW(1,\gamma,\tau_0)}(t_s) \, .
\end{align}
Thus, when fitting the curve $\log(m_k^{LW(v_L,\gamma,\tau_0)}(t_s))$ to $\log(m_k^{emp}(t_s))$, we can just estimate $v_L$ considering only one moment as
\begin{equation}
    \log{\hat{v}_L} = \frac{1}{k} \left( \langle \log{(m_k^{LW(1,\gamma,\tau_0)}(t_s))} \rangle_{t_s} - \langle \log{(m_k^{emp}(t_s))} \rangle_{t_s} \right) \, .
\end{equation}
Since we are considering second and fourth moments, we choose the estimate of $v_L$
\begin{equation}
\hat{v}_L = \exp\!\left(
       \frac{1}{2} \sum_{k=2,4} \frac{1}{k} \left( \langle \log{(m_k^{LW(1,\gamma,\tau_0)}(t_s))} \rangle_{t_s} - \langle \log{(m_k^{emp}(t_s))} \rangle_{t_s} \right ) \right)  .
\end{equation}

The only known expressions for the moments of L\'evy walks are asymptotic in the sense that it is assumed that $\tau_0/t_s\sim 0$. In this case, the scaling of the momenta with $t_s$ takes the form:
\begin{equation}
m_{2n}^{LW}(t_s) \sim
\begin{cases}
t_s^{2n}, & \gamma < 1, \\
t_s^{2n + 1 - \gamma}, & 1 < \gamma < 2, \\
t_s^{n}, & \gamma > 2.
\end{cases}
\label{eq:Levy}
\end{equation}
In this asymptotic regime, some expressions relating the moments and $\tau_0$ and $v_L$ are known for particular parameterizations of the waiting time distribution (see Eq.~(1) of the main text)~[S1]. However, in our case, we observe significant deviations from these expressions although, as we see in Figure~\ref{fig:Levy_fits}, the moments $m_{2}^{LW}(t_s)$ and $m_{4}^{LW}(t_s)$ are well approximated by a power-law (a line in a log-log plot).

\begin{figure}[b]
\begin{center}
\includegraphics[width=0.45\textwidth]{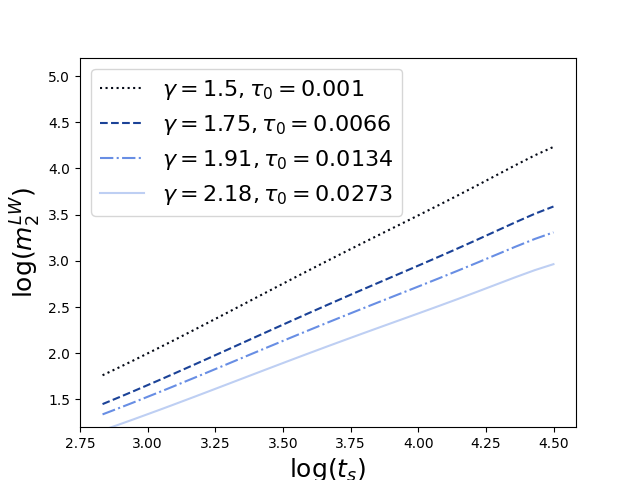}%
\includegraphics[width=0.45\textwidth]{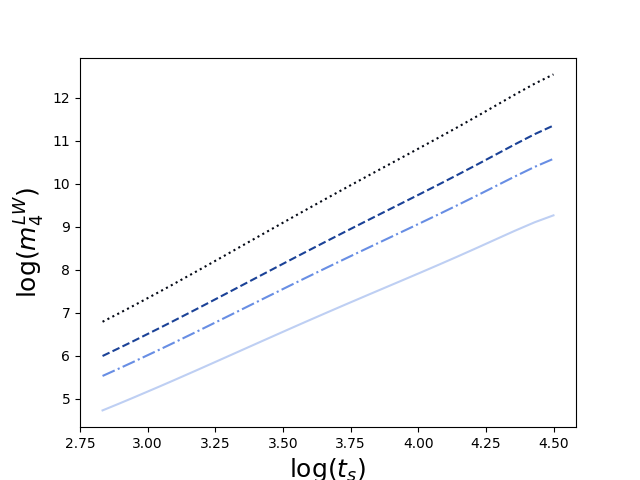}
\end{center}
\caption{\protect
Sampled values of $m_2^{LW(1,\gamma,\tau_0)}t$ and $m_4^{LW(1,\gamma,\tau_0)}(t_s)$ for different values of $\gamma$ and $\tau_0$. We observe that, although there is a slight curvature for some of the scaling of moments, the curves are well approximated by a line.}
\label{fig:Levy_fits}
\end{figure}
\begin{figure}[t]
\begin{center}
\includegraphics[width=0.3\textwidth]{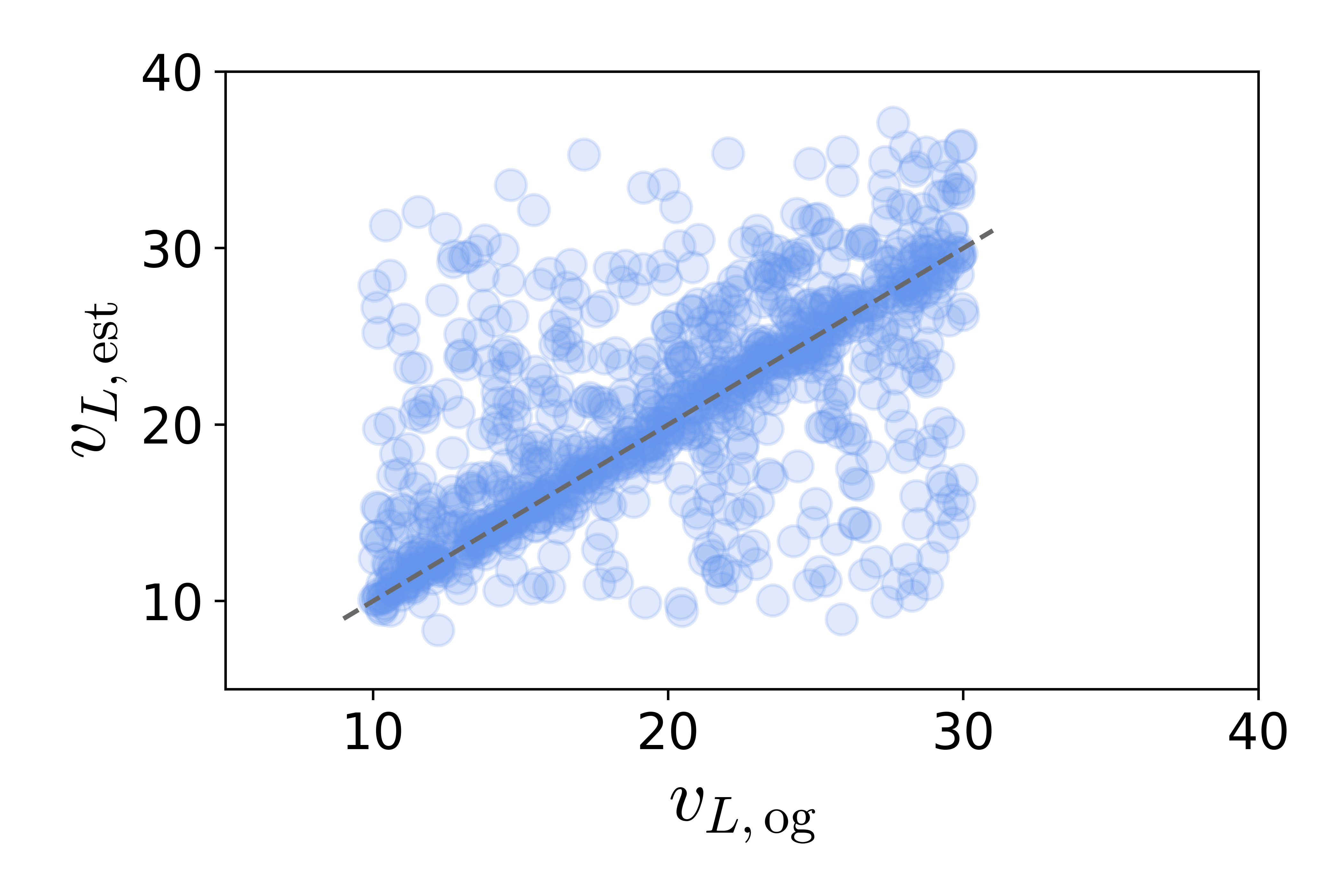}
\includegraphics[width=0.3\textwidth]{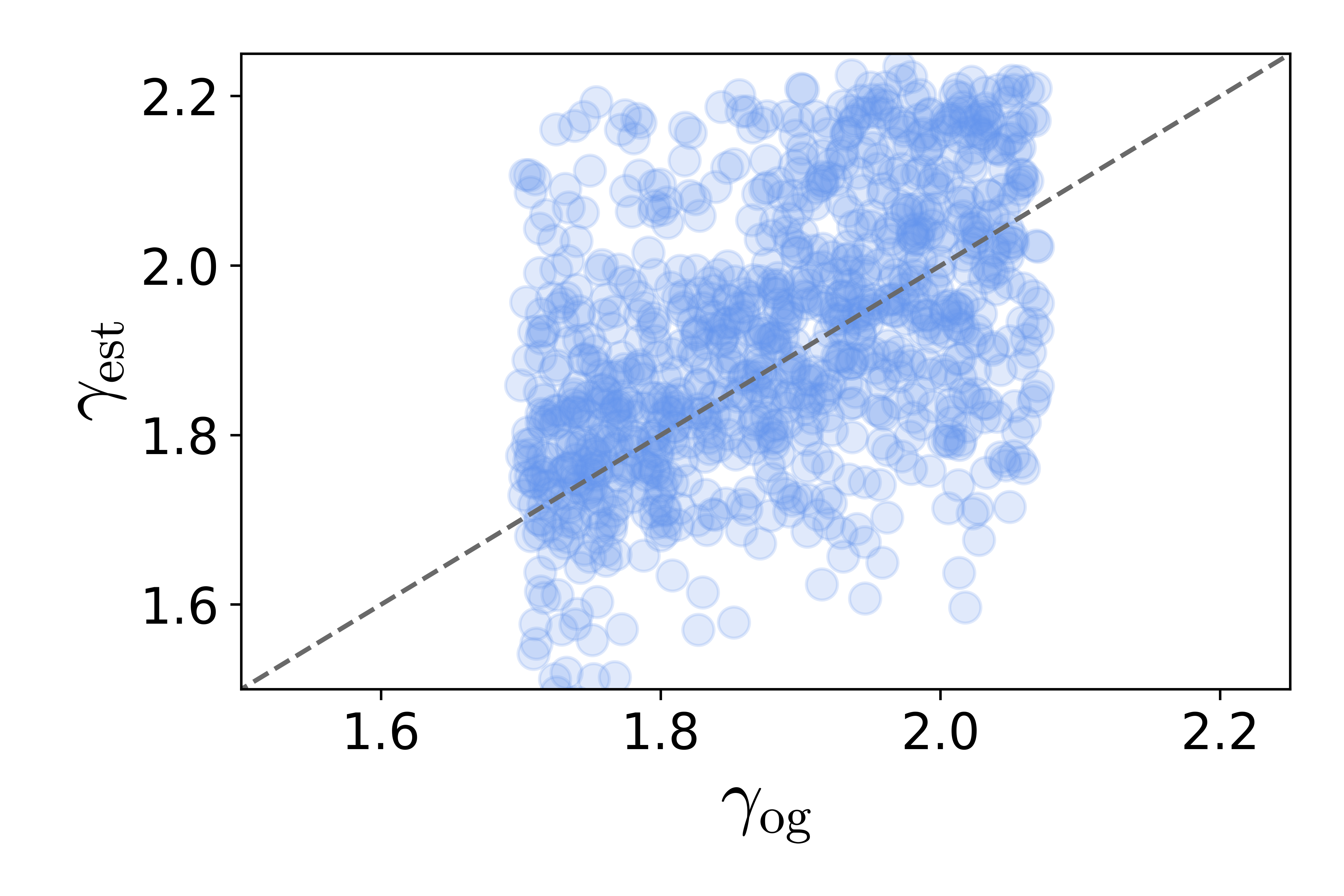}
\includegraphics[width=0.3\textwidth]{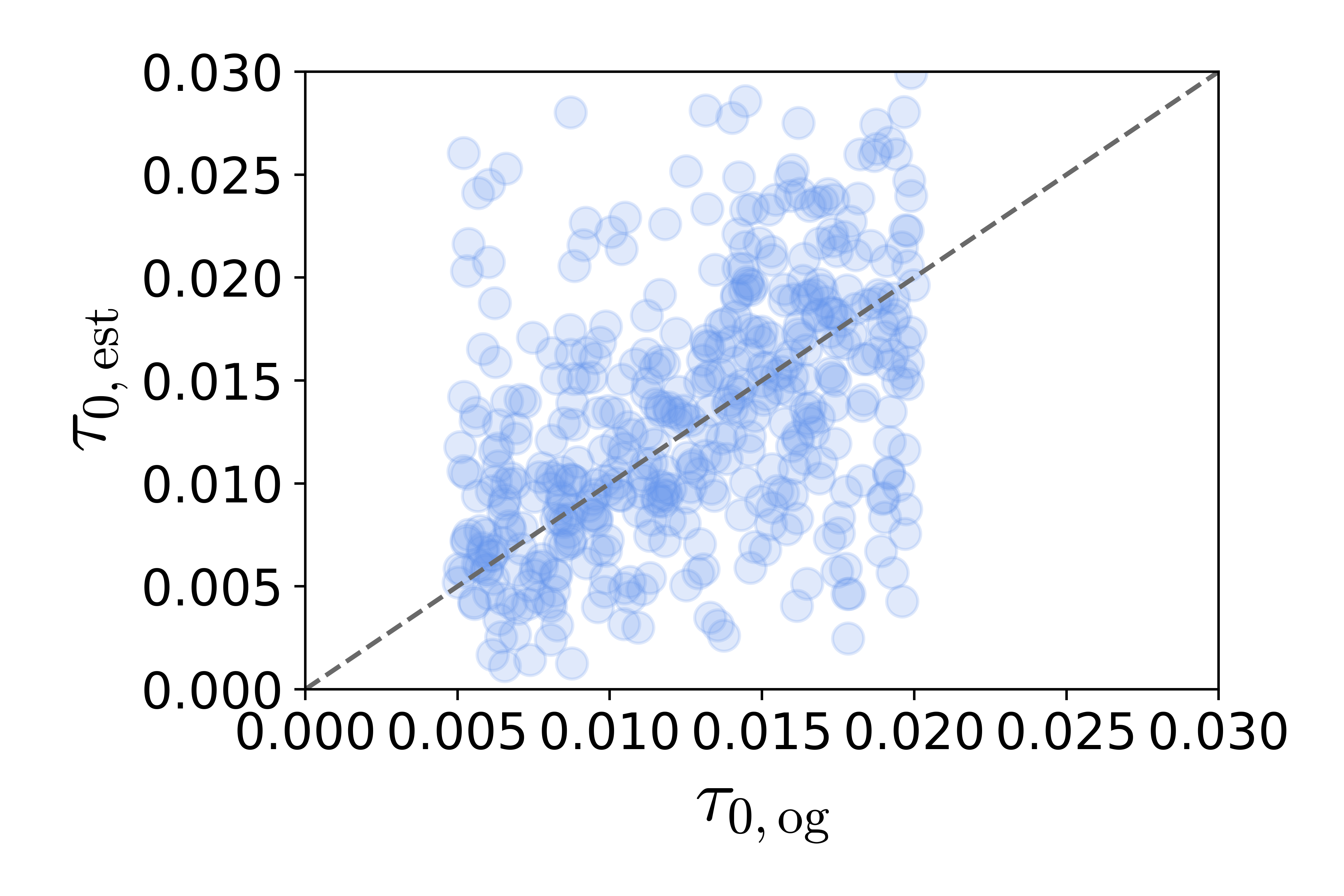}
\includegraphics[width=0.3\textwidth]{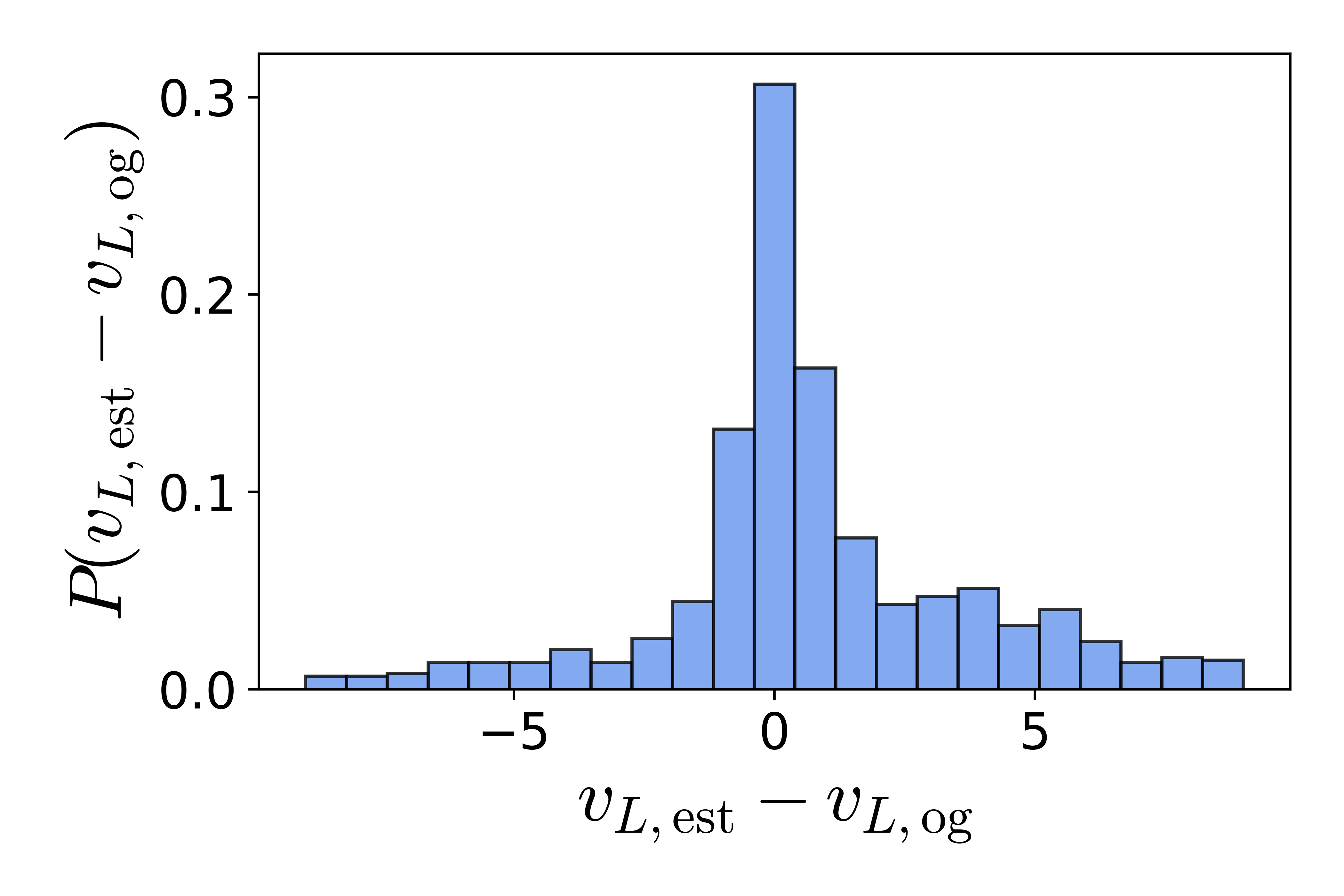}
\includegraphics[width=0.3\textwidth]{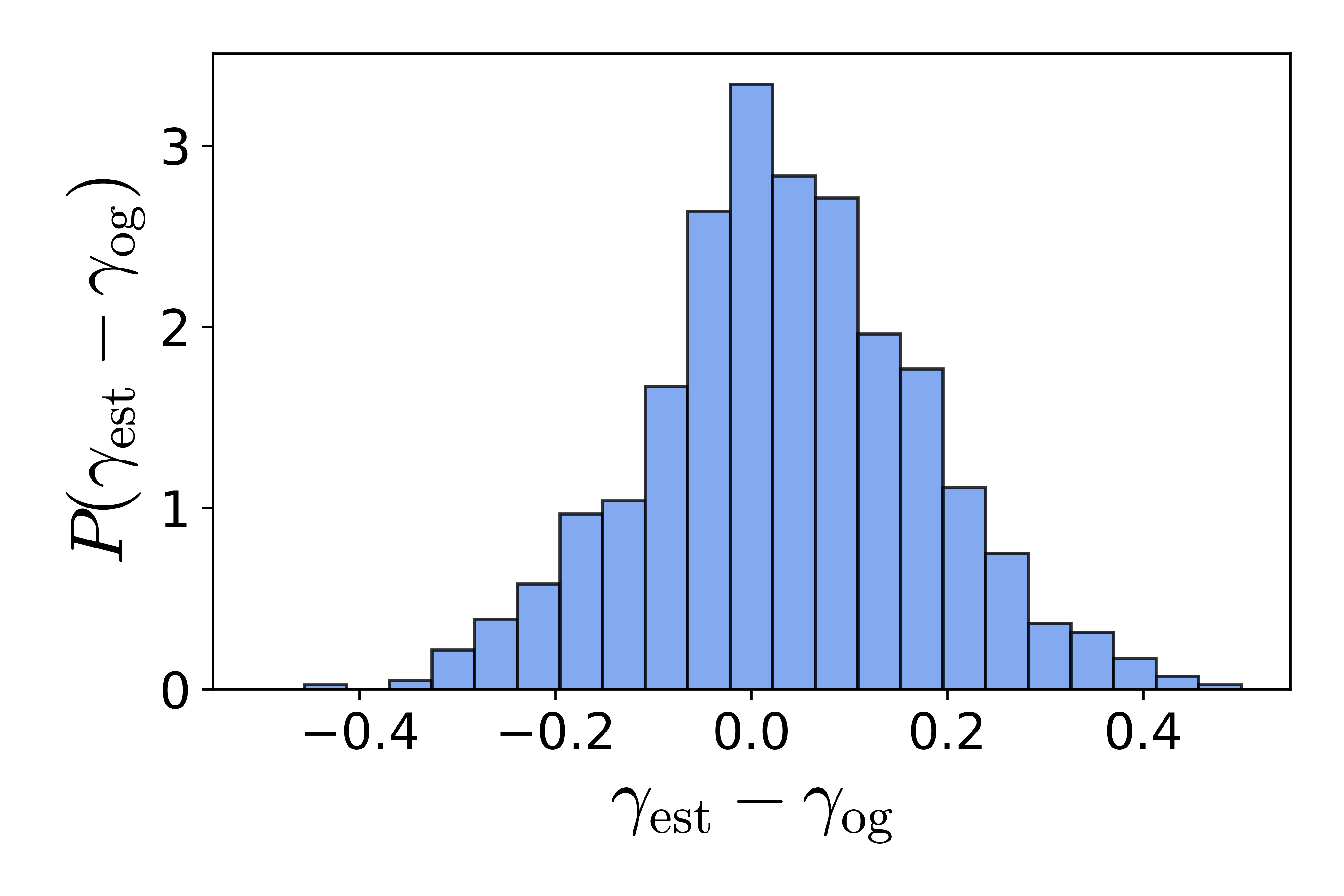}
\includegraphics[width=0.3\textwidth]{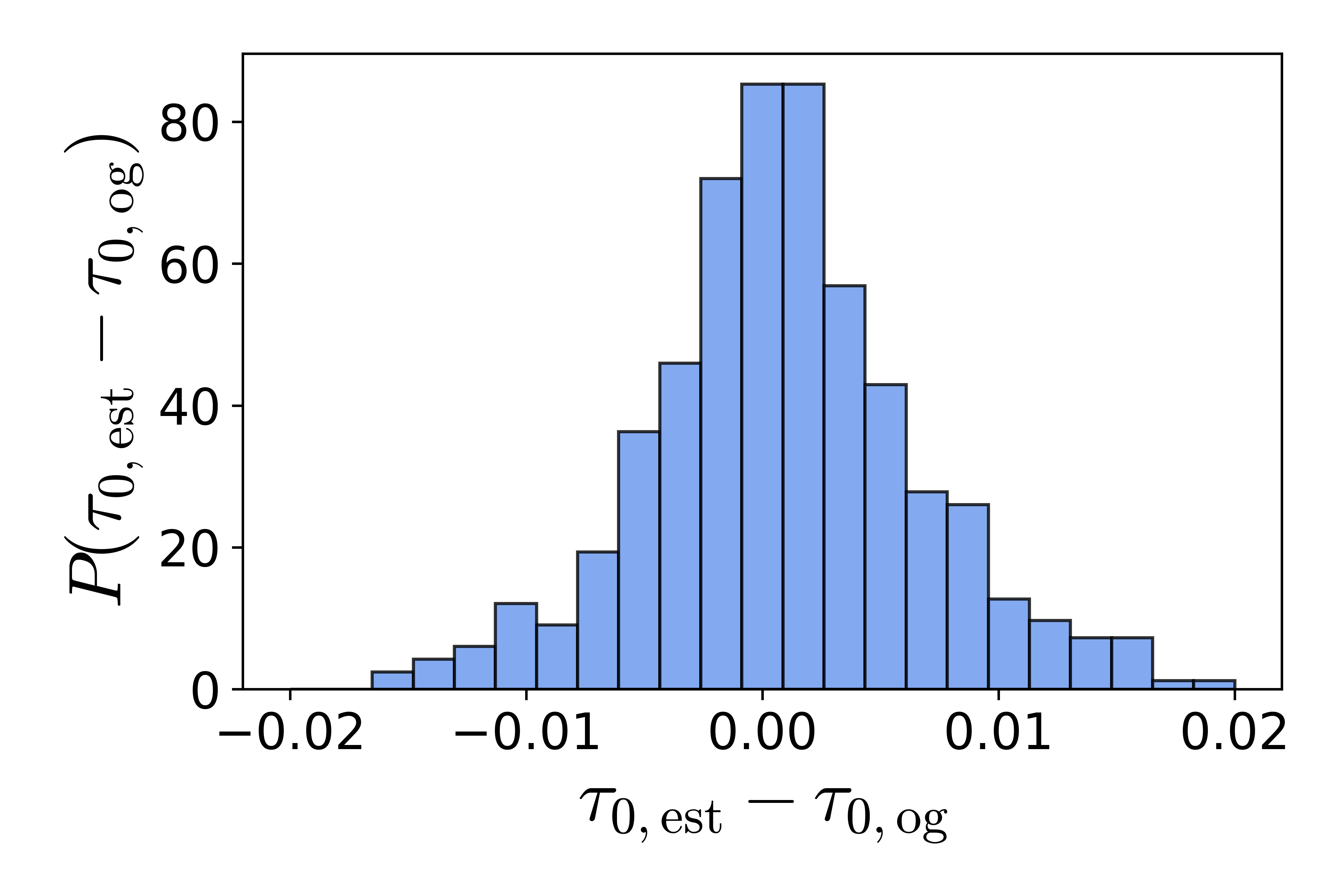}
\end{center}
\caption{\protect
Benchmark of the estimation accuracy of the simulated LW parameters. We observe that there is not a significant bias in the estimation of the parameters. However also note that there is a significant dispersion in the estimation of $\gamma$ relative to the range of values that $\gamma$ can take (cf. Figure 4), namely that the standard deviation $\sigma(\gamma_{\text{og}} - \gamma_{\text{est}}) = 0.14$, where $\gamma_{\text{og}}$ represents the original simulated parameter and $\gamma_{\text{est}}$ the estimated value. The same happens for $\tau_0$: with a range of values between $0.001$ and $0.02$ we have $\sigma(\tau_{0,\text{og}} - \tau_{0,\text{est}}) = 0.006$ }
\label{fig:lev_params}
\end{figure}
\begin{table}[t]
\begin{center}
\begin{tabular}{|c|c|c|c|c|c|c|c|}
\hline
 & $v_L$ & $\gamma$ & $\tau_0$ & $v_B$ & $D$ & $\lambda_B$ & $\lambda_D$ \\
\hline
$\rho$ & $8.23e-01$ & $-3.48e-02$ & $1.04e-03$ & $1.12$ & $-1.51e-01$ & $-3.26e-01$ & $-7.56e-03$ \\
\hline
$\sigma$ & $5.28$ & $1.40e-01$ & $5.96e-03$ & $8.77$ & $1.85e-01$ & $2.06e-01$ & $4.07e-03$ \\
\hline
\end{tabular}
\end{center}
\caption{Bias and standard deviation of the LW and IS parameters as defined in Eq.~\eqref{eq:rho}, Eq.~\eqref{eq:sigma}. We observe a relatively low bias for the LW parameters, except for $v_L$, which has some positive deviation, especially in large values. Regarding the parameters of the IS, we observe some bias in the $v_B$ and $D$ parameters. This can be corroborated by a visual inspection of Figure~\ref{fig:lev_params} and Figure~\ref{fig:int_params}.}
\label{tab:bias-variance}
\end{table}

We thus create a grid of values of $m_2^{LW(1,\gamma,\tau_0)}(t_s)$ and $m_4^{LW(1,\gamma,\tau_0)}(t_s)$ which is created by sampling $10^5$ L\'evy walks with $N=11250$ data points with a specific $\gamma$ and $\tau_0$. The number of data points in each trajectory $N$ was chosen in order to match the number of data points we collected. In our study, we sample $1000$ $\gamma$ values between the values of $1.4$ to $2.4$ and $1000$ values of $\tau_0$, logarithmically distributed from $0.1$ to $0.001$ to a total of $10^6$ possible combinations of values of ( $\gamma$ and $\tau_0$). Given a series of empirical points $(X_t,Y_t)$ with second moment $m^{emp}_2(t_s)$ and fourth moment $m^{emp}_4(t_s)$, we find the estimate of $\gamma$ and $\tau_0$ by finding the set of these values that maximizes ${\overline {R}}^{2}_{\text{LW},k}$.

When it comes to estimating the parameters, we simulated $1000$ LWs with randomly selected values of $\gamma \in [1.75,2.05]$ and $\tau_0 \in [0.005,0.02]$ and $v_L \in [10,30]$.
To determine the quality of the estimation of the parameters, we define the bias ($\rho_Z$) and the standard deviation ($\sigma_Z$) of the estimator as:
\begin{align}
    \rho_Z &= \langle Z_{est} - Z_{og} \rangle , \label{eq:rho} \\
    \sigma_Z &= \sqrt{\langle (Z_{est} - Z_{og})^2 \rangle }, \label{eq:sigma}
\end{align}
where $Z$ standards for a parameter of either the LW or IS model.

The results can be seen in Tab.~\ref{tab:bias-variance}, where, in conjunction with Figure~\ref{fig:lev_params},  we observe different dynamics when it comes to estimating the different $W$ parameters. In particular, we observe that there is a slight positive bias when it comes to $v_L$, but a very small or negligible one when it comes to $\gamma$ and $\tau_0$, while when it comes to $\sigma_Z$ we observe that $v_L$ has the lowest typical deviation among all parameters.

\subsubsection{Fitting the IS moments and estimating the parameters}

\begin{figure}
\begin{center}
\includegraphics[width=0.245\textwidth]{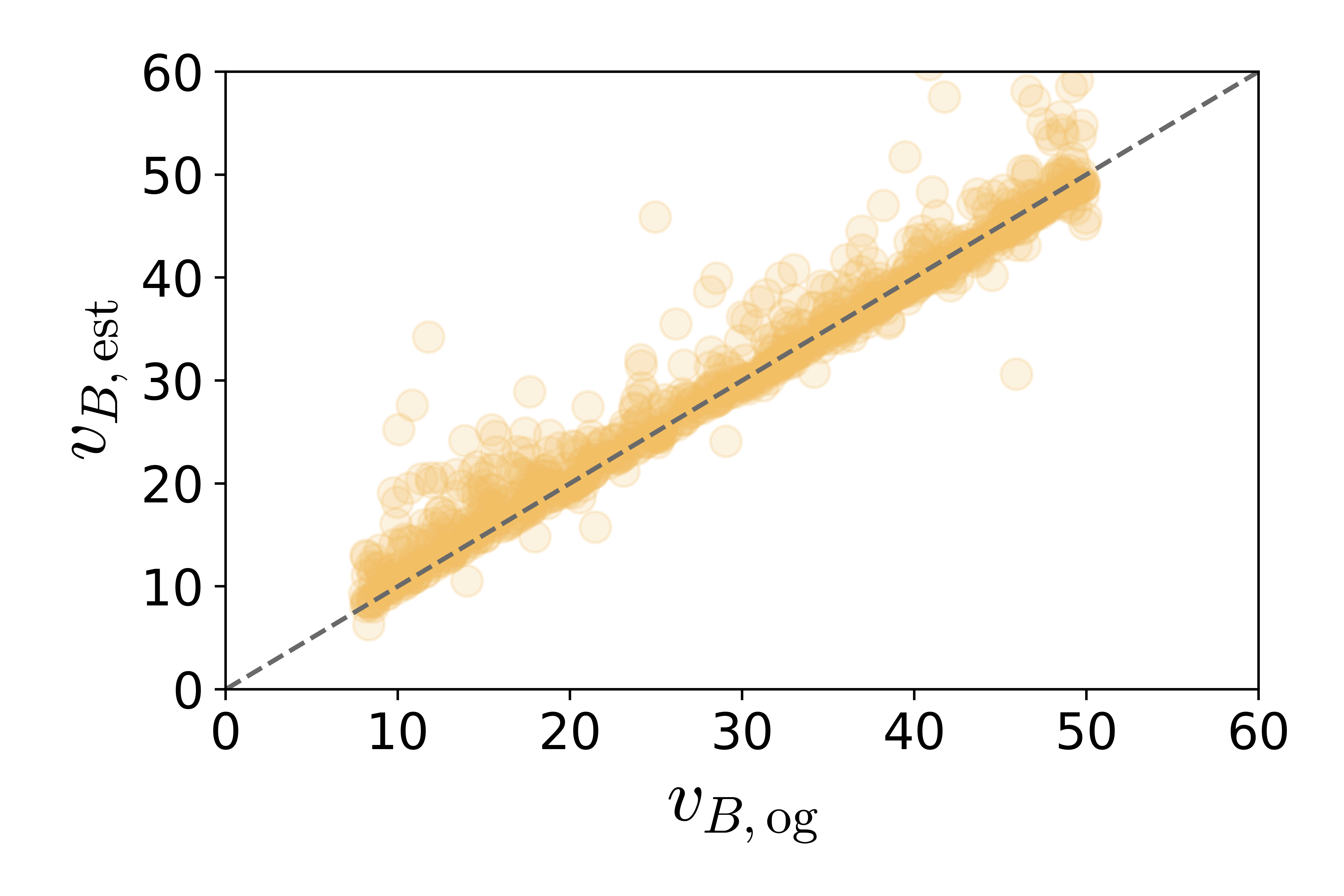}
\includegraphics[width=0.245\textwidth]{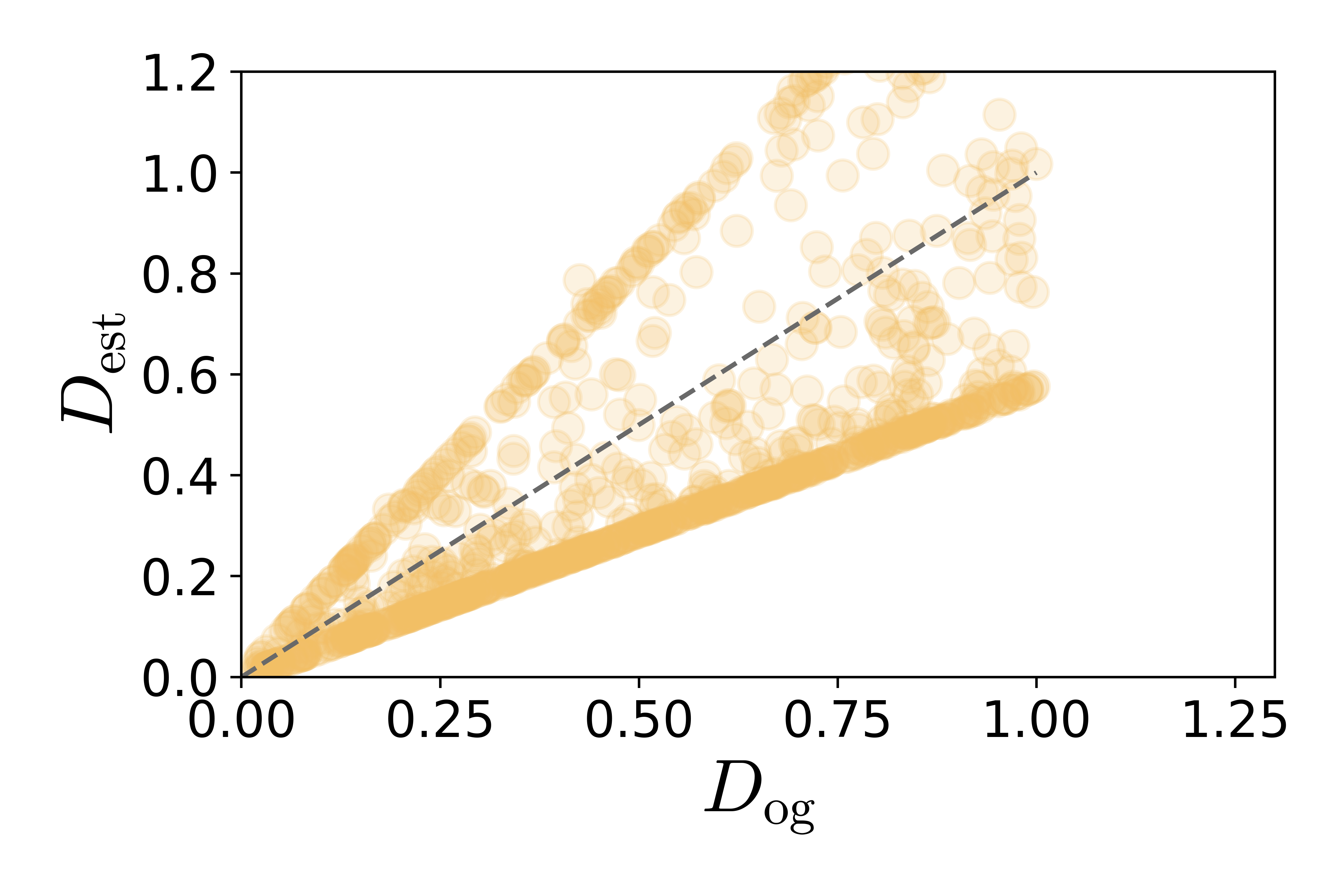}
\includegraphics[width=0.245\textwidth]{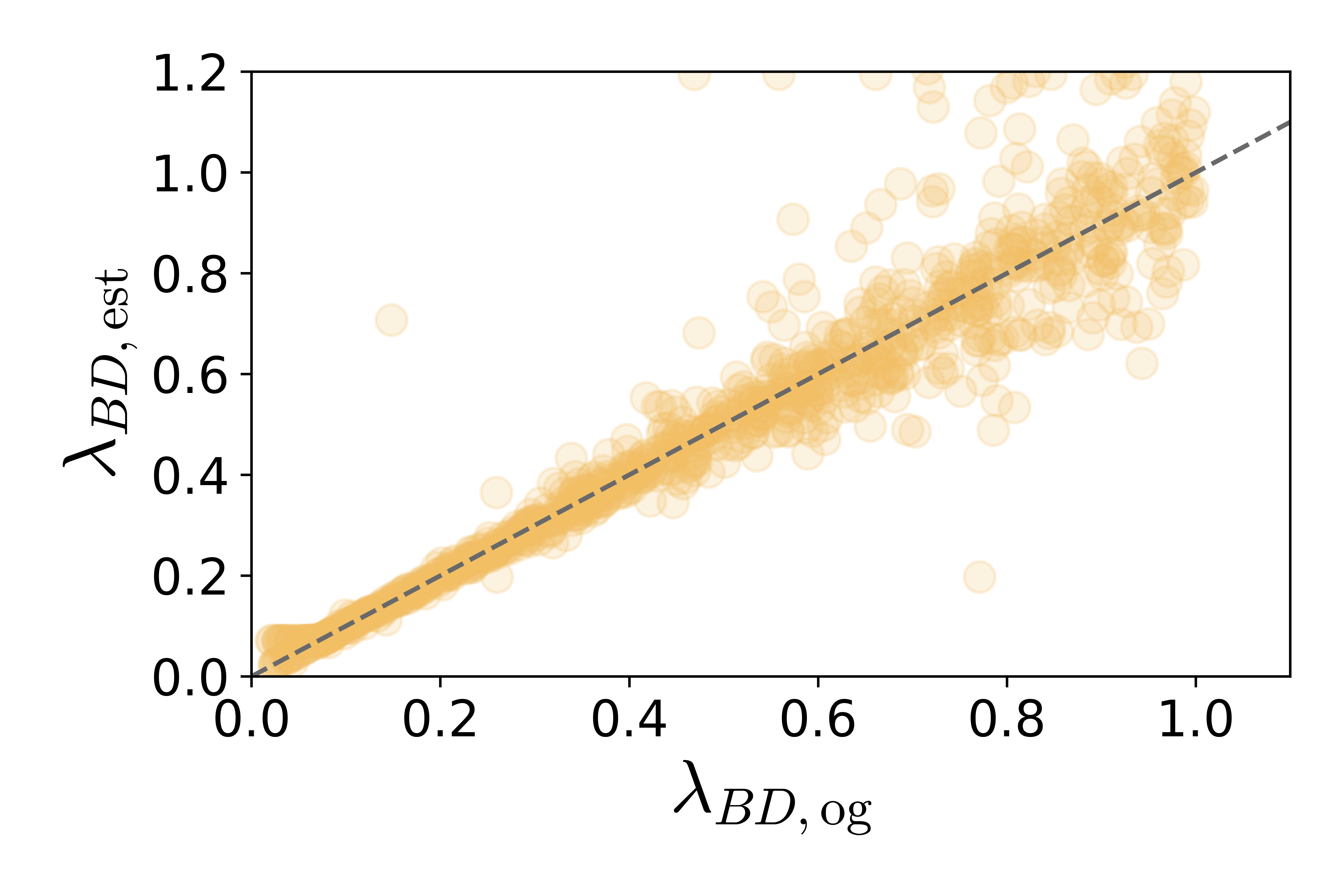}
\includegraphics[width=0.245\textwidth]{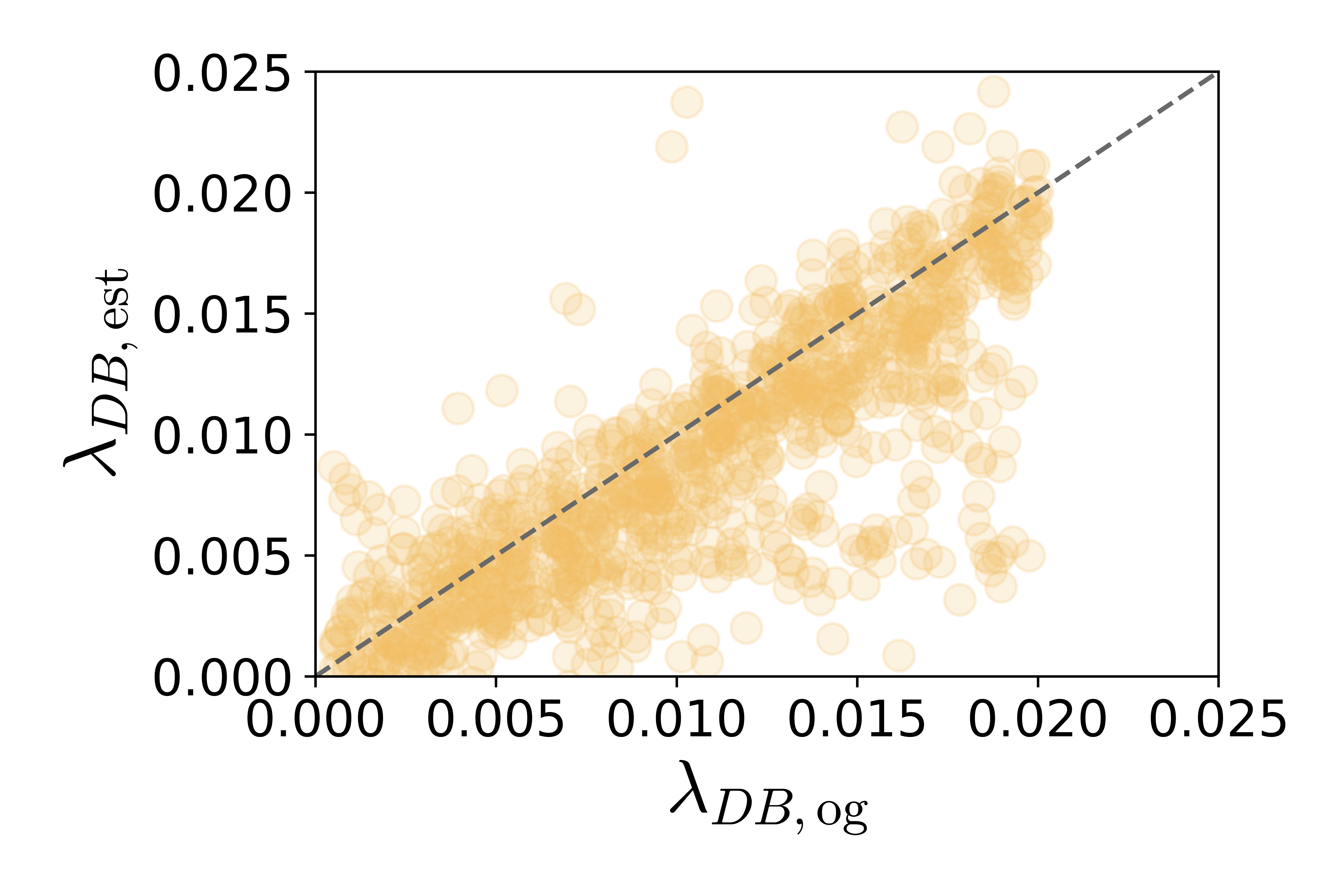}
\includegraphics[width=0.245\textwidth]{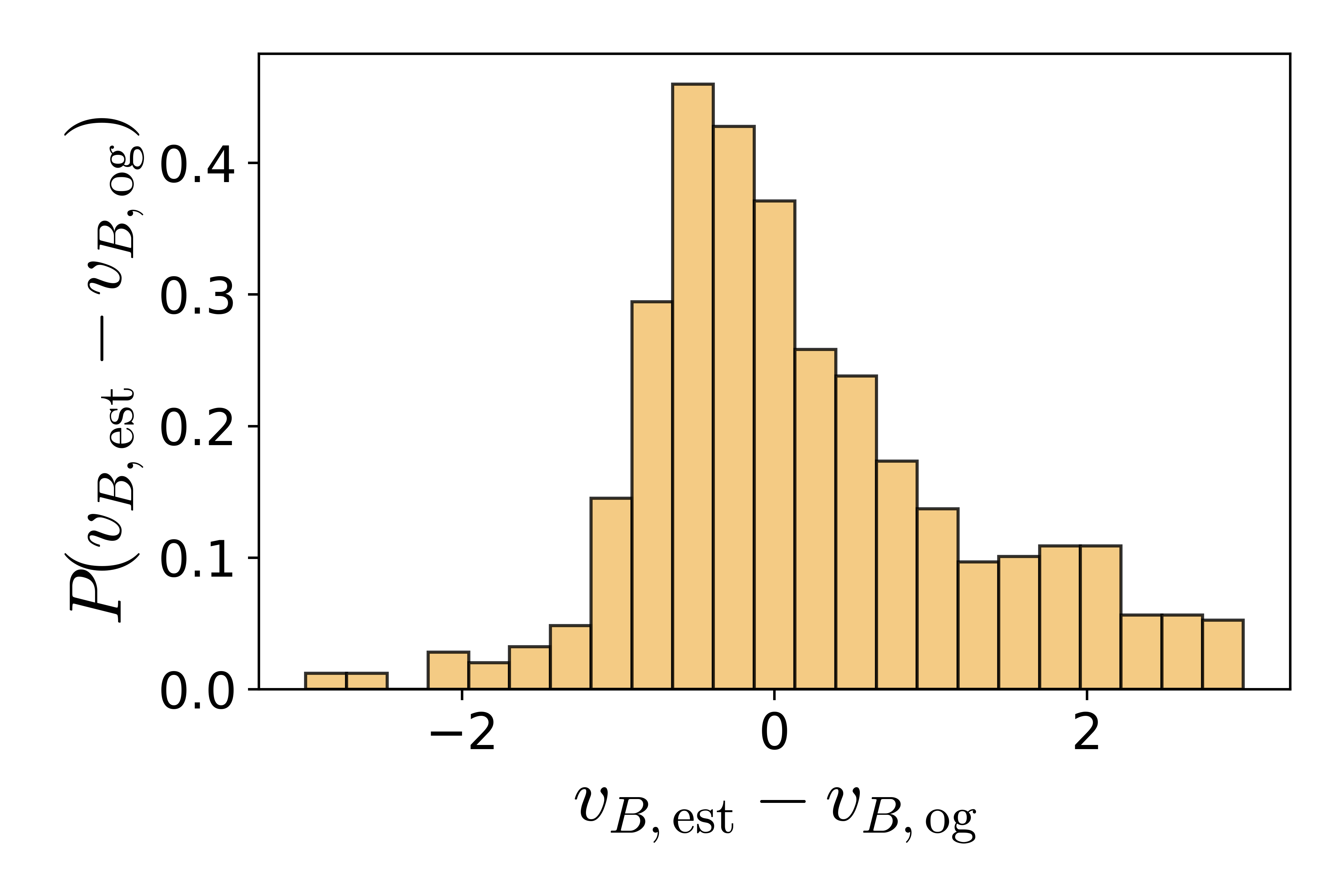}
\includegraphics[width=0.245\textwidth]{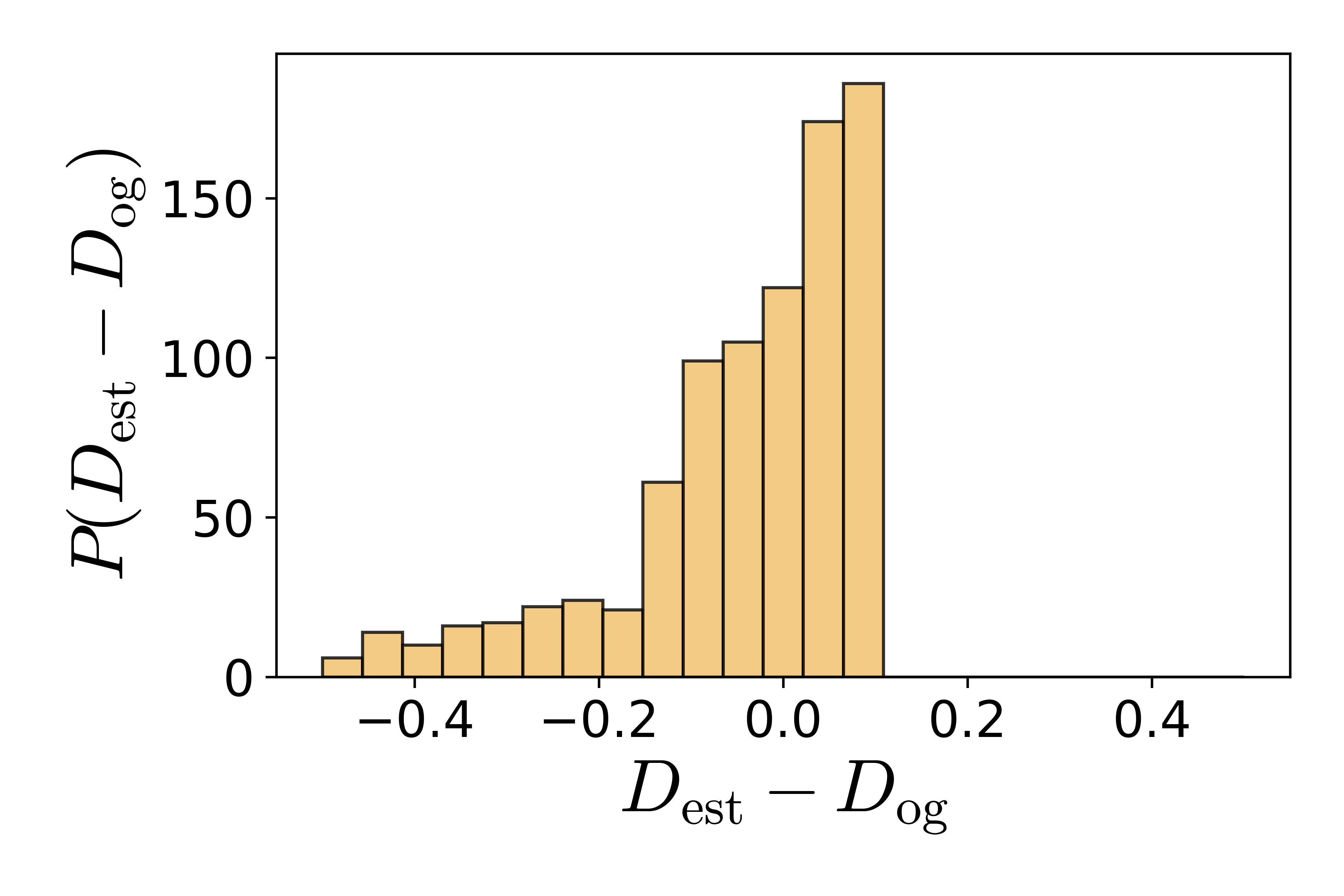}
\includegraphics[width=0.245\textwidth]{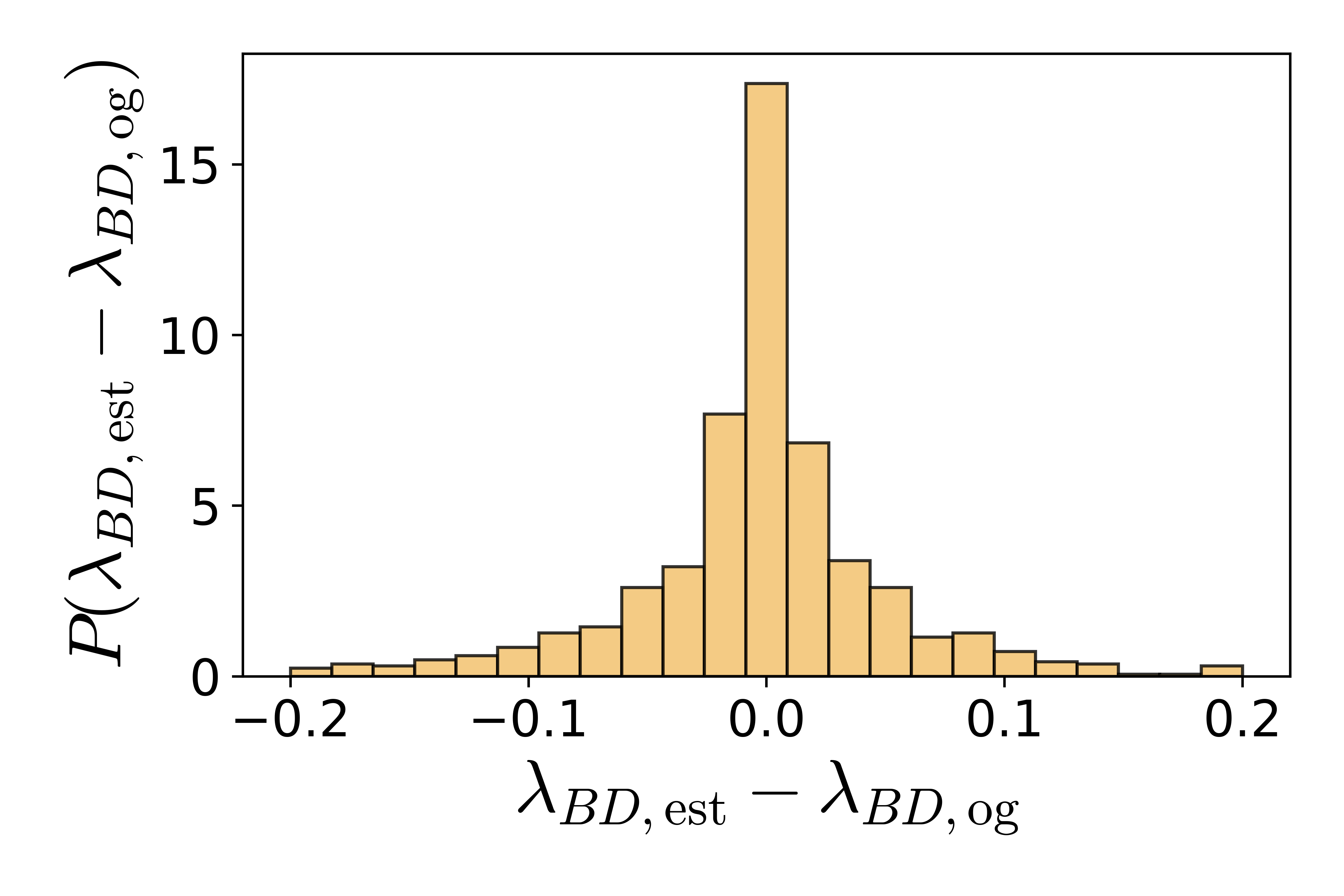}
\includegraphics[width=0.245\textwidth]{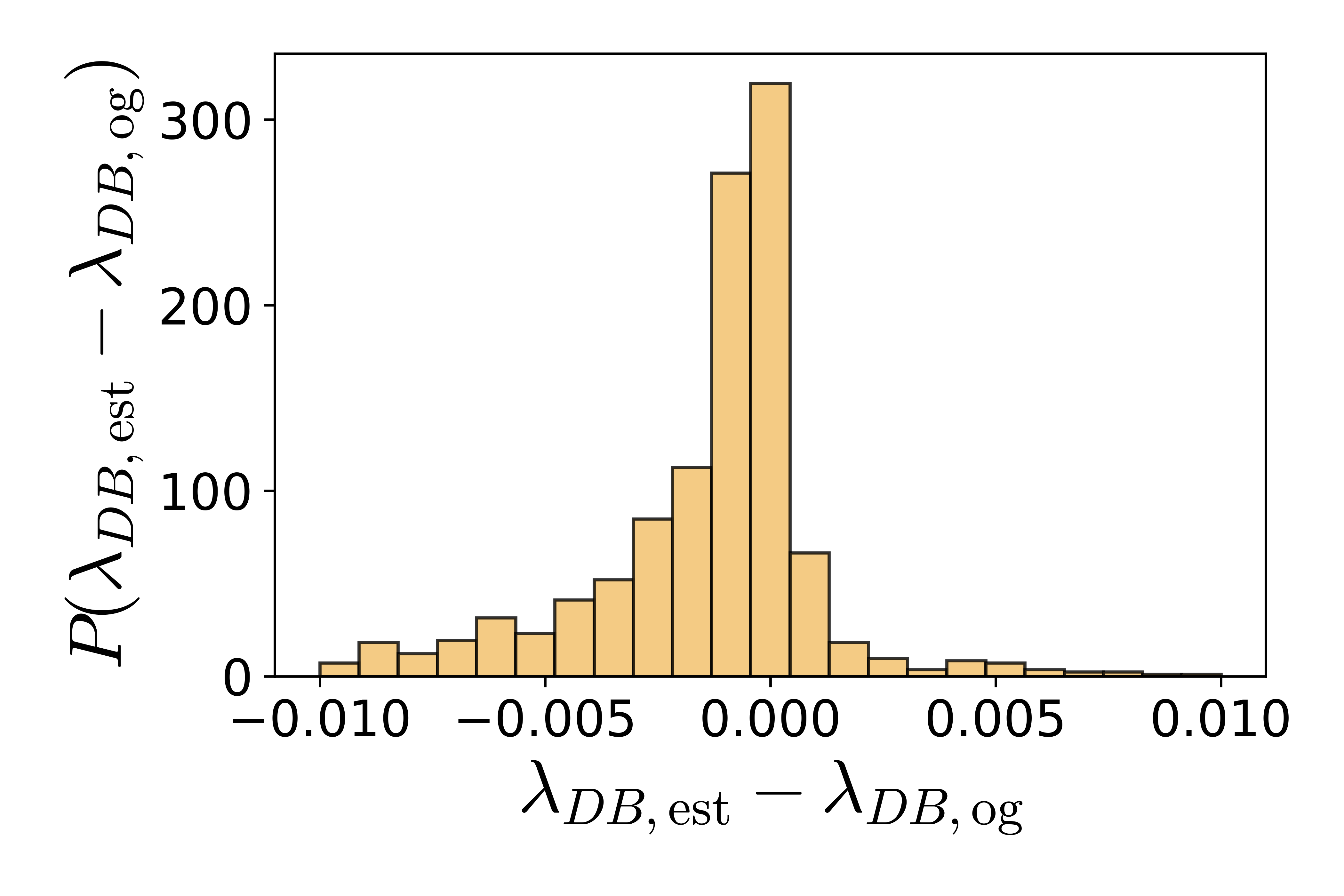}
\end{center}
\caption{\protect
Benchmark of the estimation accuracy of the simulated IS parameters. Here we observe that the parameters' estimation does not have a significant bias and that the range of fluctuations in the estimation is smaller than the range of values those parameters take in experimental settings. To note the estimation of the parameter $D$, which has a typical deviation of around $30\%$ of its simulated variable. This happens since there is a trade-off between optimizing the estimation of the parameters and obtaining the highest possible value of ${R}^{2}_{\text{IS}}$ (see text).}
\label{fig:int_params}
\end{figure}

As we mentioned before, the second moment $m_2(t_s)$ is not enough to estimate the parameters of the IS and that is the reason why $m_4(t_s)$ is needed in the first place.

However, the estimation of the IS parameters is more complex than simply optimizing for the fit of $m_4(t_s)$ or optimizing at the same time using as a target function one single function of $m_2(t_s)$ and $m_4(t_s)$. We observe that, with this approach, the estimation of $D_{\text{est}}$ typically deviates by several orders of magnitude from the value of $D_{\text{og}}$. We found, instead, that one should the parameters for $m_2(t_s)$ and $m_4(t_s)$ iteratively, as described bellow.

We also note that, given the presence of finite sampling noise, often the estimation of IS parameters finds local minima. To combat that, we will use a simulated annealing approach throughout our algorithm. Given a planar trajectory $(X_t,Y_t)$, with a second moment $m^{emp}_2(t_s)$ and fourth moment $m^{emp}_4(t_s)$ the algorithm proceeds as follows:

\begin{itemize}
    \item[1] The simulated annealing procedure optimized the parameters based on a pre-determined range of parameters. If this range is not known, as is our case, a pre-estimation of the values is performed. This is done using already established routines to estimate the parameters of hidden Markov models. This is then used to establish a rough acceptable range for the parameters.
    \item[2] After having an acceptable range of parameters, we fix the parameters $D$ and $\lambda_{DB}$ optimize the parameters $v_{B}$ and $\lambda_{BD}$ so that the expression of $\log(m_{4,IS,v_B,D,\lambda_{BD},\lambda_{DB}}(t_s))$ matches $\log(m^{\prime}_4(t_s))$ as best possible. The optimization is done via simulated annealing and several iterations each with a different starting point, which is selected from a grid of possible values of $v_{B}$ and $\lambda_{BD}$. This results in several parameter values and, from these, we chose the ones that better fit
    $ \log(m_{4,IS,v_B,D,\lambda_{BD},\lambda_{DB}}(t_s))$ to $\log(m^{\prime}_4(t_s)$ and $\log(m_{2,IS,v_B,D,\lambda_{BD},\lambda_{DB}}(t_s)))$ to $\log(m^{\prime}_2(t_s))$
    \item[3] Analogously, in this step we fix $v_{B}$ and $\lambda_{BD}$ and optimize $D$ and $\lambda_{DB}$ so that the expression of $\log(m_{2,IS,v_B,D,\lambda_{BD},\lambda_{DB}}(t_s))$ matches $\log(m^{\prime}_2(t_s))$ as best possible. Similarly, we repeat this procedure for several possible initial values.
    \item[4] Steps 2 and 3 are repeated $20$ times. Each iteration is accepted if it improves the fit of the moments. This is done until a final value of the parameters is found.
    \item[5] Even if we have used simulated annealing and taken further precautions to avoid local minima in our estimation, this still happens. For that reason, we repeat steps 2, 3, and 4 a total of 5 times. This allows us to detect if some estimation of parameter values is orders of magnitude different than the other estimations or if the moment fitting is significantly worse for some given set of parameters.
    \item[6] Up until this point we have optimized $\log(m_2(t_s))$ and $\log(m_4(t_s))$ one at a time, for a different set of parameters. In this final step we optimize all parameters and try to approximate  $ \log(m_{4,IS,v_B,D,\lambda_{BD},\lambda_{DB}}(t_s)) + 2\times \log(m_{2,IS,v_B,D,\lambda_{BD},\lambda_{DB}}(t_s))$ to $\log(m^{\prime}_4(t_s)) + 2\times \log(m^{\prime}_2(t_s))$. Usually, to have a good fit for both moments, a final adjustment of parameters is needed and, in this step, we allow the parameters to become $1.3$ times larger or smaller. Without this final step, usually, the estimation of parameters is a bit better, but the overall fit of the second and fourth moments is worse. This is also the reason the behavior of $D_{\text{est}}$ relative to $D_{\text{og}}$ seen in Figure~\ref{fig:int_params}.
\end{itemize}

In Figure~\ref{fig:int_params} we can see the dependency of the estimated parameters on the originally simulated ones. Besides the case of the diffusion coefficient $D$, which was already discussed, we see a larger deviation from the original values when it comes to $\lambda_{DB}$, while we observe a larger bias in the values of $v_{B}$ (see $\rho_{v_B}$ in Table~\ref{tab:bias-variance}). The remaining values of $\rho$ and $\sigma$ relative to the quality of the estimation of the IS parameters can be seen in Table~\ref{tab:bias-variance}.

\subsubsection{Discriminating between L\'evy and Intermittent walks}

After determining the parameters of the LW and IS, we can check which of the models is the best approximation of gaze trajectories by computing $\Gamma$ (see Eq.~(5)  of the manuscript). We simulate 1000 LW and IS trajectories with the parameter range observed in Figure~\ref{fig:lev_params} and Figure~\ref{fig:int_params} and observe that over $99\%$ of LW are well classified, with only 6 of them being labeled as IS (see Figure~\ref{fig:gamma_bench}). On the other hand, of the 1000 simulated IS, 20 were mislabeled as LW. Thus, in a similar range of parameters, as we found in our eye-tracking experiment, we observe that the $\Gamma$ classifier presents a high accuracy in determining if a given trajectory is better approximated by a LW or IS.

\begin{figure}
\begin{center}
\includegraphics[width=0.98\textwidth]{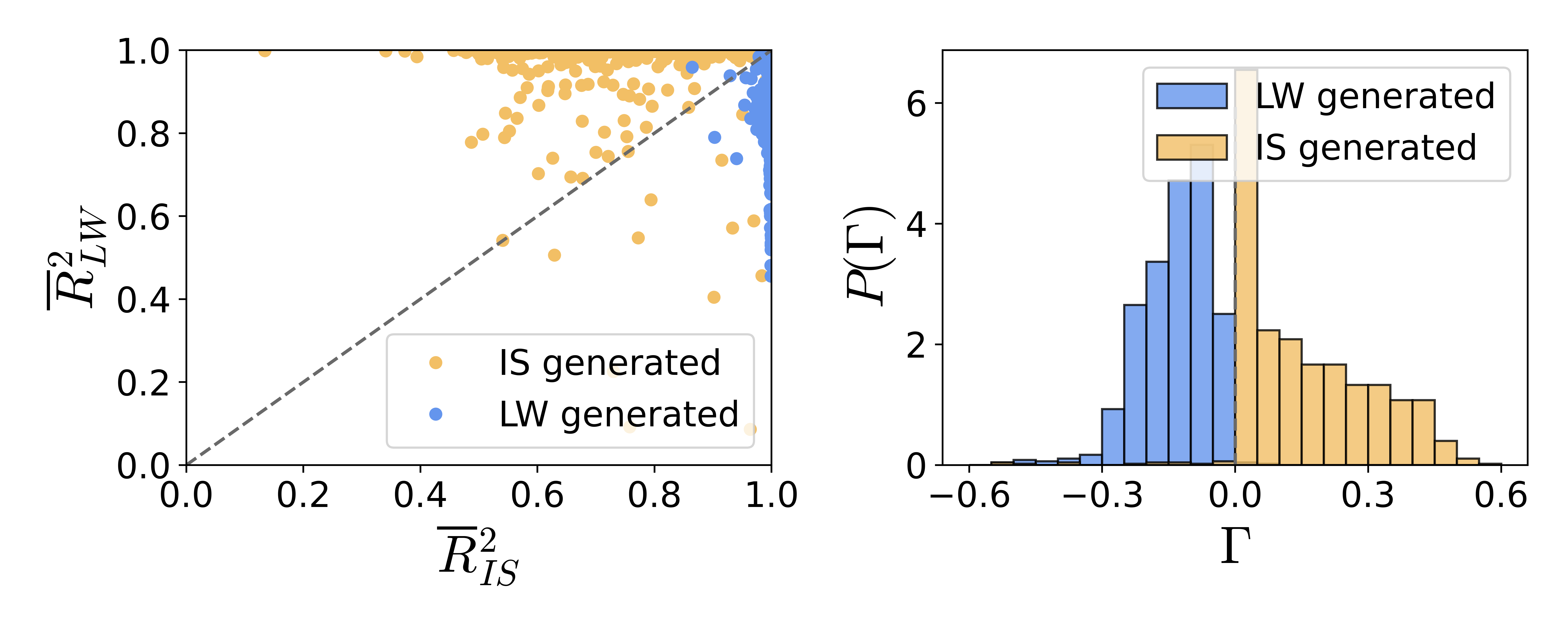}
\end{center}
\caption{\protect Benchmark of synthetically generated LW (blue) and IS trajectories (yellow). Specifically, the parameter $\gamma$ (see Eq.~(5) of the main text) is less than zero ($\gamma < 0$) for the vast majority of LW trajectories (996 out of 1000). Conversely, for IS trajectories, $\gamma$ exceeds zero ($\gamma > 0$) in 980 out of 1000 cases. The IS and LW trajectories were simulated using the same estimated parameters and number of data points as the empirical eye-tracking trajectories described in the main text (cf. Figure 4 of the main manuscript).}
\label{fig:gamma_bench}
\end{figure}



\vspace*{12pt}
\noindent\textbf{\large{{References}}}

\noindent [S3] A. Rebenshtok, S. Denisov, P. H¨ anggi, and E. Barkai,  Phys. Rev. E 90, 062135 (2014).

\end{widetext}

\end{document}